\documentclass[fleqn,usenatbib]{mnras}
\usepackage{newtxtext}

\usepackage[T1]{fontenc}
\usepackage{siunitx}

\usepackage{graphicx}	
\usepackage{amsmath}	
\usepackage{amssymb}	

\title[PHANGS Star Cluster and GMC Correlation]{PHANGS: Constraining Star Formation Timescales Using the Spatial Correlations of Star Clusters and Giant Molecular Clouds}

\author[J. A. Turner et al.]{
Jordan~A.~Turner,$^{1}$\thanks{E-mail: jturner6563@gmail.com}
Daniel~A.~Dale,$^{1}$
James~Lilly,$^{1}$
Mederic~Boquien,$^{2}$
Sinan~Deger,$^{3}$
Janice~C.~Lee,$^{4}$
\newauthor
Bradley~C.~Whitmore,$^{5}$
Gagandeep~S.~Anand,$^{5}$
Samantha~M.~Benincasa,$^{6,7}$
Frank~Bigiel,$^{8}$
\newauthor
Guillermo~A.~Blanc,$^{9,10}$
M\'elanie~Chevance,$^{11}$
Eric~Emsellem,$^{12,13}$
Christopher~M.~Faesi,$^{14}$
\newauthor
Simon~C.~O.~Glover,$^{15}$
Kathryn~Grasha,$^{16,24}$\thanks{ARC DECRA Fellow}
Annie~Hughes,$^{17,18}$
Ralf S.\ Klessen,$^{15,19}$
Kathryn~Kreckel,$^{11}$
\newauthor
J.~M.~Diederik Kruijssen,$^{11}$
Adam K. Leroy,$^{6,7}$
Hsi-An~Pan,$^{20}$
Erik Rosolowsky,$^{21}$
Andreas~Schruba,$^{22}$ 
\newauthor
Thomas~G.~Williams$^{23}$
\\
$^{1}$Department of Physics \& Astronomy, University of Wyoming, Laramie, WY 82071, USA \\
$^{2}$Universidad de Antofagasta, Centro de Astronomía, Avenida Angamos 601, Antofagasta 1270300, Chile\\
$^{3}$TAPIR, California Institute of Technology, Pasadena, CA 91125, USA\\
$^{4}$Gemini Observatory/NOIRLab, 950 N. Cherry Avenue, Tucson, AZ, 85719, USA \\
$^{5}$Space Telescope Science Institute, 3700 San Martin Drive, Baltimore, MD 21218, USA\\
$^{6}$Department of Astronomy, The Ohio State University, 140 West 18th Avenue, Columbus, Ohio 43210, USA\\
$^{7}$Center for Cosmology and Astroparticle Physics, 191 West Woodruff Avenue, Columbus, OH 43210, USA\\
$^{8}$Argelander-Institut f\"{u}r Astronomie, Universität Bonn, Auf dem H\"{u}gel 71, 53121 Bonn, Germany\\
$^{9}$The Observatories of the Carnegie Institution for Science, 813 Santa Barbara St., Pasadena, CA 91101, USA \\
$^{10}$Departmento de Astronom\'ia, Universidad de Chile, Camino del Observatorio 1515, Las Condes, Santiago, Chile \\
$^{11}$Astronomisches Rechen-Institut, Zentrum f{\"u}r Astronomie der Universit{\"a}t Heidelberg, M{\"o}nchhofstra{\ss}e 12-14, 69120 Heidelberg, Germany\\
$^{12}$European Southern Observatory, Karl-Schwarzchild Stra\ss e 2, D-85748 Garching bei M\"unchen, Germany\\
$^{13}$Univ Lyon, Univ Lyon 1, ENS de Lyon, CNRS, Centre de Recherche Astrophysique de Lyon UMR5574, F-69230 Saint-Genis-Laval, France\\
$^{14}$University of Massachusetts—Amherst, 710 N. Pleasant Street, Amherst, MA 01003, USA\\
$^{15}$Universit\"at Heidelberg, Zentrum f\"ur Astronomie, Intitut f\"ur Theoretische Astrophysik, Albert-Ueberle-Str 2, D-69120 Heidelberg, Germany\\
$^{16}$Research School of Astronomy and Astrophysics, Australian National University, Canberra, ACT 2611, Australia\\
$^{17}$CNRS, IRAP, 9 Av. du Colonel Roche, BP 44346, F-31028 Toulouse cedex 4, France\\
$^{18}$Universit\'e de Toulouse, UPS-OMP, IRAP, F-31028 Toulouse cedex 4, France\\
$^{19}$Universit\"{a}t Heidelberg, Interdisziplin\"{a}res Zentrum f\"{u}r Wissenschaftliches Rechnen, Im Neuenheimer Feld 205, D-69120 Heidelberg, Germany\\
$^{20}$Department of Physics, Tamkang University, No.151, Yingzhuan Road, Tamsui District, New Taipei City 251301, Taiwan\\
$^{21}$4-183 CCIS, University of Alberta, Edmonton, Alberta, Canada\\
$^{22}$Max-Planck-Institut f{\"u}r extraterrestrische Physik, Giessenbachstra{\ss}e 1, D-85748 Garching, Germany\\
$^{23}$Max-Planck-Institut f\"{u}r Astronomie, K\"{o}nigstuhl 17, D-69117, Heidelberg, Germany\\
$^{24}$ARC Centre of Excellence for All Sky Astrophysics in 3 Dimensions (ASTRO 3D), Australia\\
}

\date{Accepted XXX. Received YYY; in original form ZZZ}

\pubyear{2022}

\begin{document}
\label{firstpage}
\pagerange{\pageref{firstpage}--\pageref{lastpage}}
\maketitle

\begin{abstract}
In the hierarchical view of star formation, giant molecular gas clouds (GMCs) undergo fragmentation to form small-scale structures made up of stars and star clusters. Here we study the connection between young star clusters and cold gas across a range of extragalactic environments by combining the high resolution (1\arcsec) PHANGS--ALMA catalogue of GMCs with the star cluster catalogues from PHANGS--HST. The star clusters are spatially matched with the GMCs across a sample of $11$~nearby star-forming galaxies with a range of galactic environments (centres, bars, spiral arms, etc.).  We find that after $4{-}6$~Myr the star clusters are no longer associated with any gas clouds. Additionally, we measure the autocorrelation of the star clusters and GMCs as well as their cross-correlation to quantify the fractal nature of hierarchical star formation. Young (${\leq}10$~Myr) star clusters are more strongly autocorrelated on kpc and smaller spatial scales than the $>$$10$~Myr stellar populations, indicating that the hierarchical structure dissolves over time.

\end{abstract}




\section{Introduction}
Galaxies serve as the stellar factories of the universe, churning out stars formed from the gravitational collapse of the densest regions of molecular clouds inside the interstellar medium. This process depletes a galaxy of gas, which can be replenished through galactic mergers or by the infall of external gas from the circumgalactic medium. In turn, star formation is a catalyst of galaxy evolution by injecting metals, energy, and momentum back into the interstellar medium and intergalactic medium, while regulating galaxy growth by exhausting the supply of gas. Star formation may shape a galaxy's evolution, but what triggers the formation of stars? There are many physical processes that can create the dense molecular cloud environment favorable for star formation \citep{maclow04,mckee07,dobbs14}. On scales of kiloparsecs and larger, there is a measured correlation between the available gas reservoir and the rate at which a galaxy forms stars \citep[the Kennicutt--Schmidt relation;][]{schmidt59,kennicutt98,kennicutt12}. However, on small-scales, this relation breaks down with a disconnect between the physical locations of the star clusters formed and the remaining molecular gas clouds \citep{onodera10,schruba10,kruijssenlongmore14,boquien14,pessa21}. This breakdown is a direct probe of the cloud-scale physics of star formation \citep{chevance20b}.

Zooming in from a global view of star formation, we can begin to break down star formation into individual constituents and processes. The cold molecular gas is hierarchically structured and at the peak of the hierarchy, i.e., the densest regions, the star clusters are formed from fragmentation and collapse of the cold, molecular gas \citep{elmegreenandfalgarone96,elmegreen08,kruijssen12}. The star cluster population therefore inherits the hierarchical distribution from their natal GMCs or giant molecular clouds \citep[e.g.,][]{lada03,grasha18,grasha19,ward20}. These small-scale structures---young stars and clusters--- represent the top of the stellar hierarchy which includes large structures like associations and cluster complexes \citep{zhang01,gouliermis10,sanchez10,gouliermis18,menon21}. The structure is likely a consequence of the hierarchical nature of turbulence throughout the interstellar medium \citep{elmegreenandefremov96,hopkins2013}. The hierarchical structure of the stellar distributions (and the interstellar medium in general) can be described with a power-law, hence it is scale-free \citep{grasha18,grasha19}. Over time, this inherited hierarchical distribution is dissipated as found by \cite{gieles08} and \citet{bastian09lmc} for the stellar populations of the Small Magellanic Cloud and Large Magellanic Cloud, respectively, which lost their structured distributions on the timescale of each galaxy's crossing time.  

Recent observations and data sets like those from the PAWS \mbox{CO(1--0)} survey of M51 \citep{pety13,schinnerer13} and now the PHANGS--ALMA \mbox{CO(2--1)} survey \citep{leroy21a} allow for studying the hierarchical structure of star formation at the scales of individual molecular clouds. Combining these ${\sim}1\arcsec$ CO maps with high-resolution (${\sim}0.05\arcsec$) \textit{Hubble Space Telescope} (\textit{HST}) observations of young star clusters like from LEGUS \citep{adamo17} and PHANGS--HST \citep{lee22} provides an unprecedented look at the connection between individual clouds and the products of star formation. The quantification of how star clusters and molecular gas are structured across spatial scales is still in its infancy. Recently, \citet{grasha19} combined the PAWS CO map of M51 with the LEGUS star clusters and found that after $6$~Myr the star clusters were no longer associated with their natal GMCs. Additionally, the authors quantified the hierarchical distribution of the clusters and molecular gas using angular two-point correlation functions. This powerful analysis afforded by the high resolution data allows us to study how the stars and gas are organized in multi-scale structures, how those structures evolve in time, and measure some of the timescales of star formation---how long do star clusters stay associated with their natal gas clouds? 

\citet{kruijssenlongmore14} and \citet{kruijssen18} introduced a method to statistically correlate star-forming regions within galaxies from small to large spatial scales. This ``uncertainty principle for star formation'' empirically measures star formation efficiencies and the timescales for gas clouds to collapse, form stars, and provide stellar feedback without the need for individual gas clouds to be resolved, and has been successfully applied to a sample of nearby star-forming galaxies \citep{kruijssen19,ward20b,zabel20,chevance20,kim21}. These studies have provided detailed measurements of the timescales involved in star formation. \citet{chevance20} find the molecular cloud lifetime to be short at $10{-}30$~Myr and star formation efficiencies are measured ranging from a few to 10~per~cent.  The duration of the embedded phase of star formation lasts $2{-}7$~Myr \citep{kim21}, and after the onset of star formation, the parent gas clouds are dispersed within $1{-}5$~Myr which suggests that early or pre-supernova stellar feedback (like stellar winds and photoionization) are a driving factor in the dispersal of the gas clouds \citep{chevance20c}. 

In this study, we follow similar methodologies as \citet{grasha18,grasha19} to constrain the timescales of star formation as well as the evolution of the structured distributions of gas and star clusters. We utilize the PHANGS--ALMA GMC catalogs \citep[][Hughes et al., in preparation]{rosolowsky21} generated from the PHANGS--ALMA \mbox{CO(2--1)} data \citep{leroy21a,leroy21b} along with the star cluster catalogs from PHANGS--HST \citep{thilker21,deger22,lee22}. By connecting the star clusters with their natal gas clouds, we are able to provide an independent measurement of the cloud dispersal timescale which can be compared to the results from \citet{chevance20,chevance20c}.  

In Section~\ref{sec:data}, we discuss the two data sets and the 11~galaxies used in this study. In Section~\ref{sec:anl}, we explain in detail the analyses---a nearest-neighbour analysis, how we associate star clusters with GMCs, angular two-point correlation functions, and angular cross-correlation functions---using the galaxy NGC~1566 as an example. We then present the results for all 11 galaxies in our sample and discuss the results in Section~\ref{sec:rd}. We conclude with a summary in Section~\ref{sec:conc}.

\section{Data}
\label{sec:data}

The PHANGS--ALMA survey is a large \mbox{CO(2--1)} mapping program of 90~nearby galaxies with resolutions at GMC size scales of ${\sim}100$~pc.  The sample selection and properties of the galaxies are described in \citet{leroy21a} and the data processing and pipeline are described in \citet{leroy21b}. In addition to \mbox{CO(2--1)} maps, PHANGS--ALMA produces a catalogue of GMCs in each galaxy using the methods detailed by \citet{rosolowsky21} and Hughes et al. (in preparation). The GMC catalogs include cloud positions, velocities, radii, masses, and luminosities. For this study, we use the `native resolution' GMC catalogs which have been constructed with the best available resolution and noise for each galaxy. This leads to heterogeneous limits across the sample \citep[see e.g.,][]{rosolowsky21}. 

The cloud radii are measured as an average of the deconvolved major and minor axes of the elliptical profile as output by CPROPS, a decomposition algorithm for identifying GMCs in molecular-line observations \citep{rosolowsky06,rosolowsky21}. This assumes a spherically symmetric cloud which can become inaccurate if the cloud sizes approach the scale height of the molecular gas disk in the galaxy. Therefore, the clouds are modeled to be oblate spheroidal if the spherical radius exceeds the assumed scale height ($100$~pc for the PHANGS--ALMA sample). Across the ensemble of galaxies in our sample, the mean GMC radius is 61.5~pc, median is 62~pc, minimum is 8~pc and maximum is 132~pc.

The PHANGS--HST star cluster catalogs provide a robust sample of compact star clusters across 38~nearby spiral galaxies within the PHANGS--ALMA sample \citep{lee22}. Star clusters are identified and classified in the five-band ($NUV$-$U$-$B$-$V$-$I$) \textit{HST} images \citep{thilker21} and aperture photometry is performed \citep{deger22}. Spectral energy distribution fitting provides the cluster ages, masses, and dust reddening, with respective 1$\sigma$ uncertainties of ${\sim}0.3$~dex, $0.2$~dex, and $0.1$~mag for the PHANGS--HST pilot study focused on NGC~3351 \citep{turner21}; systematics include uncertainties in the photometric flux calibration (${\sim}5$\%), incomplete/incorrect priors, and SED templates. The compact star cluster catalogs thus include cluster positions, aperture photometry, ages, stellar masses, and reddening, as well as visual classifications (for a subset of the sample), neural network morphological classifications \citep{wei20,whitmore21}, and a variety of concentration index values that indicate the difference in magnitudes for different aperture radii \citep{thilker21,deger22}. For this paper, we focus on star clusters which have been visually classified as either Class~1 (symmetric and compact), Class~2 (asymmetric and compact), or Class~3 (multi-peaked compact association). The Class~3 compact associations, which are typically young, prove to be difficult to model given their potential for containing multiple ages within a single association, which causes large uncertainties in their age measurements. In order to better capture the young stellar populations, Larson et al.\ (in preparation) have developed a method to select stellar associations using a watershed algorithm. PHANGS--HST provides a stellar association catalogue similar to the compact cluster catalogue \citep{lee22}.

\begin{figure*}
    \centering
    \includegraphics[width=\textwidth]{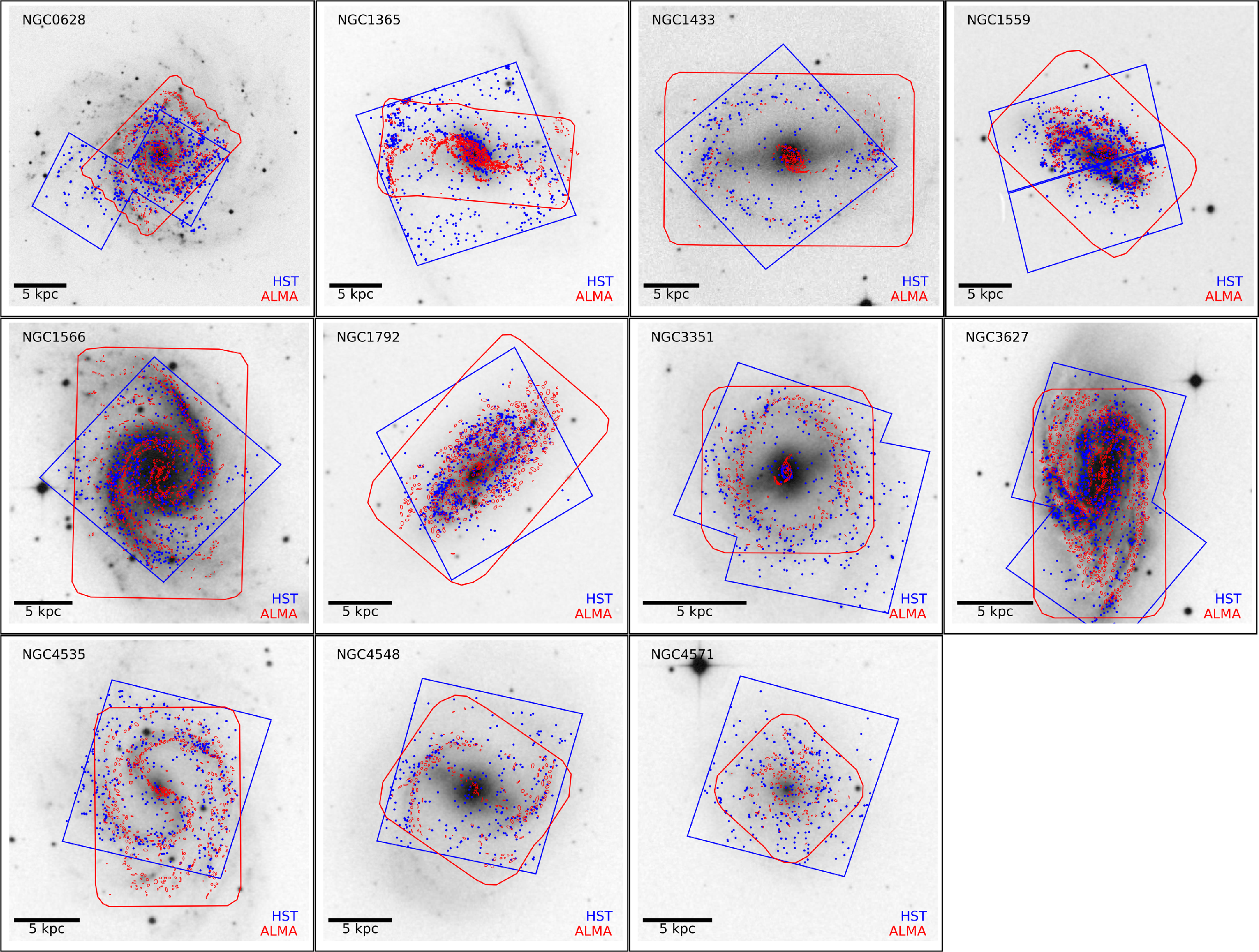}
    \caption{The PHANGS--ALMA footprints and the size and location of GMCs are shown in red. PHANGS--HST F336W observation footprints and the location of the Class~1, 2, and~3 star clusters are shown in blue. Background images are from the Digitized Sky Survey.}
    \label{fig:outline}
\end{figure*}

Table~\ref{tab:galaxies} gives an overview of the galaxy sample used in this study and Figure~\ref{fig:outline} shows the spatial distribution of the star clusters and GMCs within each galaxy. As the development of the PHANGS--HST cluster catalogue and PHANGS--ALMA GMC catalogue pipelines are still on-going, we focus on 11~PHANGS--HST galaxies which currently have completed cluster and GMC catalogs. 

\begin{table*}
\centerline{Table 1. Galaxy Sample}
\begin{tabular}{cccrrlrcrc}
\hline\hline
Galaxy  & RA      & DEC     & Distance & $R_{25}$ & Optical & SFR & log M$_{*}$ & \# of & \# of star\\
        & [J2000] & [J2000] & [Mpc]    & [kpc]    & Morph.  & [$M_{\odot}$ yr$^{-1}$] & [log $M_{\odot}$] & GMCs   & clusters \\
\hline
NGC~0628 & 01h36m41.75s & $+$15d47m01.2s &  9.84 & 14.1  & SAc    &  1.7 & 10.34              & 811    & 789 \\ 
NGC~1365 & 03h33m36.37s & $-$36d08m25.4s & 19.57 & 34.2  & SBb    & 17.4 & 11.00              & 1091   & 789 \\ 
NGC~1433 & 03h42m01.55s & $-$47d13m19.5s & 18.63 & 16.8  & SBab   &  1.1 & 10.87              & 356    & 293 \\ 
NGC~1559 & 04h17m35.77s & $-$62d47m01.2s & 19.44 & 11.8  & SBcd   &  4.0 & 10.37              & 725    & 927 \\ 
NGC~1566 & 04h20m00.42s & $-$54d56m16.1s & 17.69 & 18.6  & SABbc  &  4.6 & 10.79              & 1127   & 851 \\ 
NGC~1792 & 05h05m14.45s & $-$37d58m50.7s & 16.20 & 13.1  & SAbc   &  3.7 & 10.62              & 533    & 675 \\ 
NGC~3351 & 10h43m57.70s & $+$11d42m13.7s &  9.96 & 10.5  & SBb    &  1.3 & 10.37              & 369    & 468 \\ 
NGC~3627 & 11h20m14.96s & $+$12d59m29.5s & 11.32 & 16.9  & SABb   &  3.9 & 10.84              & 984    & 958 \\ 
NGC~4535 & 12h34m20.31s & $+$08d11m51.9s & 15.77 & 18.7  & SABc   &  2.2 & 10.54              & 640    & 452 \\ 
NGC~4548 & 12h35m26.45s & $+$14d29m46.8s & 16.22 & 13.1  & SBb    &  0.5 & 10.70              & 236    & 271 \\ 
NGC~4571 & 12h36m56.38s & $+$14d13m02.5s & 14.90 &  7.7  & SAd    &  0.3 & 10.10              & 214    & 262 \\ 
\hline
\end{tabular}
\caption{Galaxy coordinates are from \citet{lee22}, $R_{25}$ sizes, star formation rates, and stellar masses are from \citet{leroy21a}, and the distances are from \citet{anand21}. Optical morphologies from \citet{dale17}. The number of star clusters include Classes 1, 2, and 3.  }
\label{tab:galaxies}
\end{table*}

For reference, the PHANGS--HST stellar cluster catalogs go much deeper than the PHANGS--ALMA GMC catalogs, at least in an absolute mass sense.  \cite{rosolowsky21} show that the 50\% completeness limit, where 50\% of mock GMCs injected into signal-free regions are recovered, is $4.7\times10^5~M_\odot$ for the PHANGS--ALMA datacubes that are homogenized to uniform noise properties and to a common 90~pc resolution.  By comparison, the PHANGS--HST stellar cluster mass limit is ${\sim}10^3 {-} 10^4~M_\odot$ depending on cluster age \citep{thilker21}, but a thorough description of the PHANGS--HST cluster sample completeness will be provided in a future paper.  However, an informed comparison of these two mass limits requires additional information.  Analysis of PHANGS--ALMA star formation and molecular cloud time scales shows that the star formation efficiency per star formation event is $4{-}10$\%, and the star formation efficiency per molecular cloud free fall time is lower by a factor of a few \citep{chevance20}.  In other words, any ultimate comparison of the relative depths of the stellar cluster and GMC catalogs must keep in mind the few percent efficiencies in converting molecular gas mass to stars.
 
\section{Analysis}
\label{sec:anl}

In this section, we detail the methodology of the analyses employed in this study. NGC~1566 is used as an example in this section while Section~\ref{sec:rd} gives the results for all 11~galaxies and discusses the results. 

\subsection{Associating Star Clusters with GMCs}
\label{subsec:anl:sc_gmc}
\begin{figure}
    \includegraphics[width=\columnwidth]{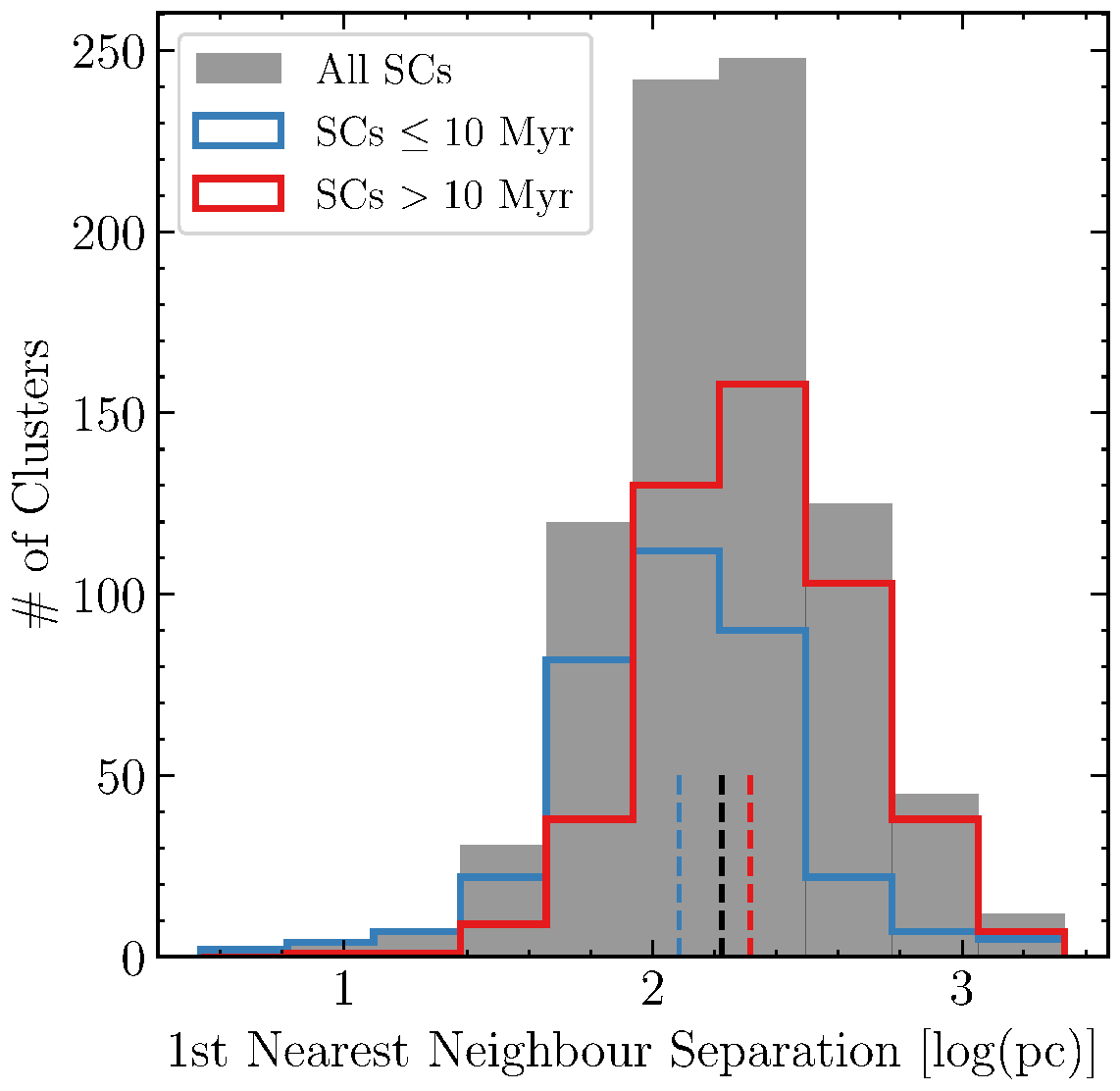}
    \caption{Histograms of the nearest neighbour separations for star clusters in NGC~1566. The star cluster sample is split into two age bins---$10$~Myr and younger (blue) and older than $10$~Myr (red). The distribution for all the star clusters is in grey. Dashed lines mark the median nearest neighbour separations.}
    \label{fig:1nnhist}
\end{figure}

In order to accurately match the star clusters with their nearest neighbour and/or natal GMC, we first find the region of overlap between the \textit{HST} and ALMA footprints and only consider clusters and clouds which lie within this overlap region. Then, as a first step in correlating the star clusters and GMCs, we perform a nearest neighbour analysis to find the separations between star clusters and the nearest GMC centres. In this analysis, the sizes of the GMCs are not taken into account. We then split the clusters into two age bins---$10$~Myr and younger and older than $10$~Myr. Figure~\ref{fig:1nnhist} shows the distribution of the first nearest neighbour separations for NGC~1566. We also test the second and third nearest neighbour separations (not shown) and find the same trend. 

Next, we search for star clusters potentially associated with GMCs by looking for line-of-sight overlap of clusters and clouds. Consistent with \citet{grasha18} and \citet{grasha19}, in this analysis the radius of a GMC is considered which allows for a cluster to be classified as either: within the radius of the closest GMC; within 1 to 2 radii of the closest GMC; within 2 to 3 radii of the closest GMC; or beyond which we consider to be ``unassociated.''  These radial bins provide a rough indication of the relative distance a star cluster has traveled since birth or the extent to which they have cleared away their parent cloud's molecular material. We only allow a star cluster to be associated with a single GMC; individual GMCs can still have multiple star clusters associated them. In cases where a star cluster is aligned with multiple GMCs, we opt for the most massive GMC to be the associated one. In terms of the typical physical extent of the GMCs in this sample of 11 galaxies, the median radius is 60.5~pc with a $16^{\rm th} {-} 84^{\rm th}$ percentile range of $40.6{-}79.7$~pc.

\begin{figure}
    \includegraphics[width=\columnwidth]{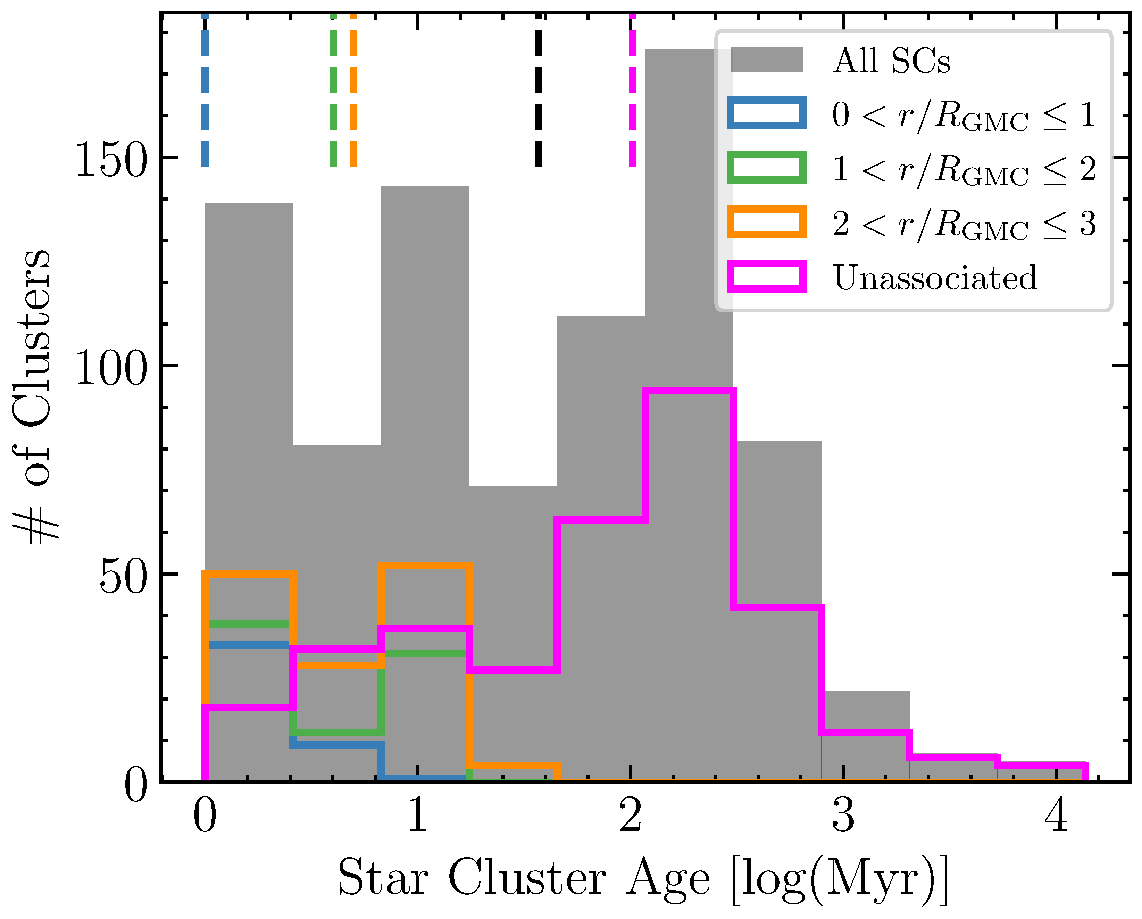}
    \caption{Histograms of the star cluster ages sorted according to association with GMCs in NGC~1566: all star clusters (grey), within one GMC radius (blue), between 1 and 2 GMC radii (green), between 2 and 3 GMC radii (orange), and beyond 3 GMC radii which is considered unassociated (fuchsia). Vertical dashed lines mark the median cluster age for each distribution. A $10$~km~s$^{-1}$ velocity cutoff is applied to minimize counting clusters which have random alignment with the GMCs.}
    \label{fig:assoc_hist}
\end{figure}

We then look for trends in cluster ages as a function of spatial association with GMCs. The distributions of the cluster ages for NGC~1566 are shown in Figure~\ref{fig:assoc_hist}.  With the distributions of the cluster ages seen in Figure~\ref{fig:assoc_hist}, we also calculate the median ages for the distributions as a means for tracking the difference in cluster ages given their association with a GMC. We calculate the uncertainty on the medians via bootstrapping the cluster ages included in the PHANGS--HST catalog.  Analysis of these trends for the full sample is provided in \S\ref{subsec:rd:nn}.

\subsection{Angular Two-Point Correlation Functions}
\label{subsec:anl:tpcf}
For the next step of the analysis, we study the distribution of both the star clusters and the GMCs with two-point correlation functions or autocorrelation functions. Autocorrelation functions help to quantify the excess probability of spatial clustering of the star formation components over a random, uniform distribution. \citet{peebles80} first applied two-point correlation functions in a cosmological context to statistically measure the clustering of mass in the large-scale structure of the universe. For such a case, the amplitude of clustering as a function of scale, $\xi(r)$, is defined to measure the excess probability above a random Poisson distribution of finding a galaxy--galaxy pair in the volume~$dV$ at a separation~$r$. In other words, if a galaxy is chosen at random from the full sample, the probability of finding a neighbouring galaxy at a distance~$r$ within the volume~$dV$ is 
\begin{equation}\label{eq:peebledP}
    dP = \bar{n} \left[1 + \xi(r)\right] dV
\end{equation}
where $\bar{n}$ is the mean number density of the galaxy sample. In this case, the autocorrelation functions are defined over a 3D~volume since the galaxy redshifts are known. On smaller scales, autocorrelation functions have been used to measure the spatial distribution of pre-main sequence stars within the Milky Way \citep{gomez93,larson95}, the distribution of resolved stellar populations in the Large Magellanic Cloud and Small Magellanic Cloud \citep{bastian09lmc, gouliermis14} and NGC~6503 \citep{gouliermis15}, and the distribution of stellar clusters in a sample of local star-forming galaxies \citep{grasha15,grasha17,grasha19}. In these cases, only a 2D~autocorrelation function is needed which can be achieved by deprojecting the positions of the stars or clusters onto a 2D~surface (the plane of the galaxy disk) given the inclination of the galaxy. Our sample of galaxies has moderate to low disk inclinations ($\lesssim60$\degr), and thus correcting for disk inclination has minimal impact and does not change our conclusions. Equation~\eqref{eq:peebledP} can be redefined to be angular so that, if an object is chosen at random from the full sample, the probability of finding a neighbouring object at an angular separation~$\theta$ within the solid angle~$d\Omega$ is
\begin{equation} \label{eq:angdP}
    dP = N \left[1 + \omega(\theta)\right] d\Omega
\end{equation}
where $N$ is the mean surface density of the sample and $\omega(\theta)$ is the amplitude of the clustering as a function of angular scale. Note that using the mean surface density implies that any large-scale gradient in the distribution of molecular gas or stars will manifest as an anti-correlation.  The expected or mean number of neighbours within the angular separation~$\theta$ of a randomly chosen object is
\begin{equation}
    \langle N \rangle_{\mathrm{p}} = N \int_{0}^{\theta} \left[1 + \omega(\theta)\right] d\Omega.
\end{equation}
With this definition, a random distribution will give $1 + \omega(\theta) = 1$ and a clustered distribution will give $1+ \omega(\theta) > 1$. An anti-correlated distribution will give $1 + \omega(\theta) < 1$.

\begin{figure}
    \includegraphics[width=\columnwidth]{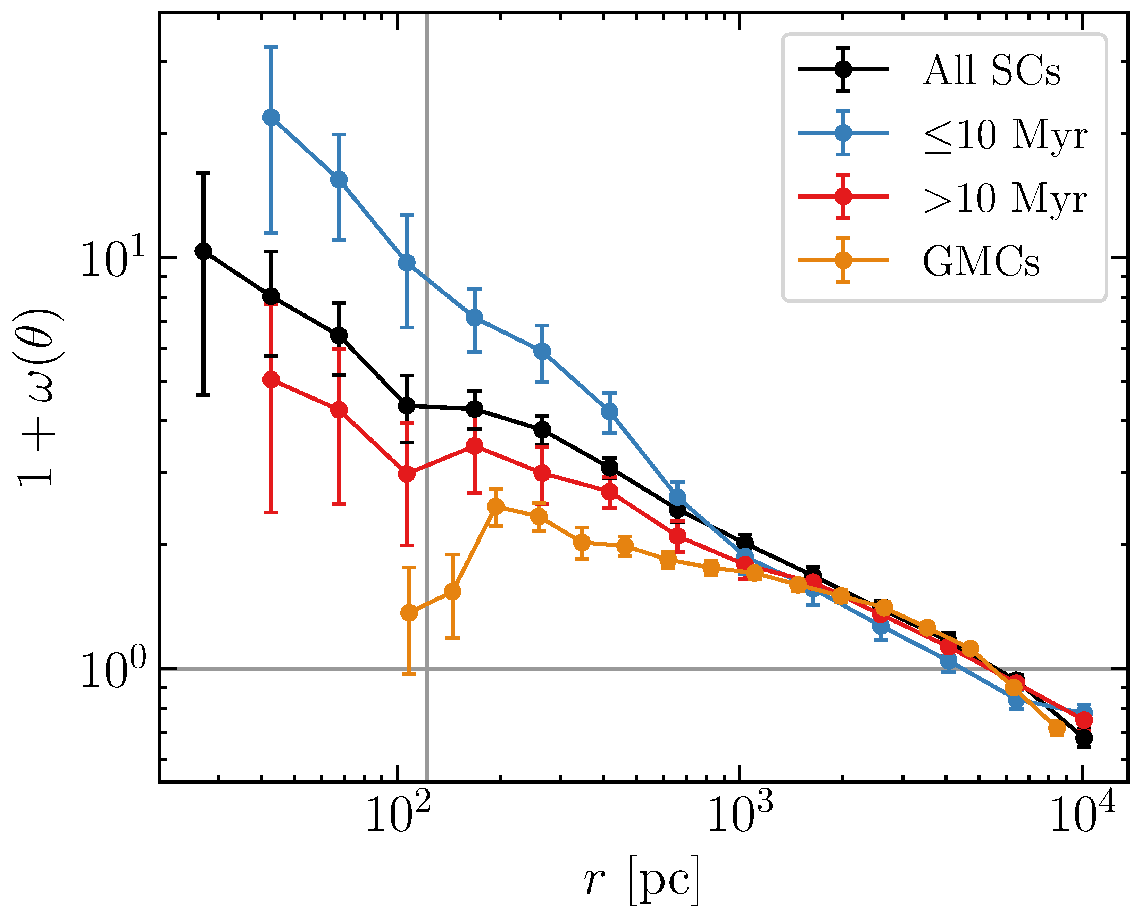}
    \caption{Angular two-point autocorrelation function $1+\omega(\theta)$ as a function of spatial scale for NGC~1566. The function for all the star clusters is black, for clusters $10$~Myr and younger is blue, for clusters older than $10$~Myr is red, and for GMCs is orange. The horizontal grey line marks a uniform, random distribution at $1+\omega(\theta) = 1$. The vertical grey line marks the median radius of the GMCs of NGC~1566. Uncertainties are bootstrap estimates.}
    \label{fig:tpcf}
\end{figure}

The clustering amplitude $\omega(\theta)$ is measured by taking a catalogue of real objects (e.g., the star clusters) and a catalogue of randomly positioned objects, and then counting the number of pairs of real objects (DD), number of pairs in the random catalogue (RR), and the number of pairs with one real object and one random catalogue object (DR) with separations within some angular separation bin. The number of pairs are normalized by the total number in that catalogue so that 
\begin{align}
    {\rm DD} &= \frac{\text{number of real object pairs}}{N_{\rm D} N_{\rm D}},\\
    {\rm RR} &= \frac{\text{number of random catalogue pairs}}{N_{\rm R} N_{\rm R}},\\
    {\rm DR} &= \frac{\text{number of real-random catalogue pairs}}{N_{\rm D} N_{\rm R}},
\end{align}
where $N_{\rm D}$ is the total number of objects in the real catalogue and $N_{\rm R}$ is the total number of objects in the random catalog. $\omega(\theta)$ is then estimated using the \citet{landy93} estimator of the form
\begin{equation}
    \omega(\theta) = \frac{1}{{\rm RR}(\theta)}\left[{\rm DD}(\theta)\left(\frac{N_{\rm R}}{N_{\rm D}}\right)^{2} - 2{\rm DR}(\theta) \left(\frac{N_{\rm R}}{N_{\rm D}}\right) + {\rm RR}(\theta)\right].
\end{equation}
The estimation of $\omega(\theta)$ is dependent on the quality of the random catalogue and how it emulates any data sampling effects in the observations.  One potential issue is the impact of crowding on identifying and classifying the star clusters in the HST imaging.  However, the cluster candidate selection process demands a stringent signal-to-noise cut of 10 \citep{thilker21} and \citet{whitmore21} explains that PHANGS--HST clusters are consistently classified to within 70\% even for the most crowded regions.  The characteristics of the GMC catalogue are also unlikely to significantly skew the estimation of $\omega(\theta)$: the noise of the ALMA cubes varies only mildly ($<20$\%) over the spatial and spectral axes within a galaxy, and crowding of the GMCs is not a concern \citep{leroy21b,rosolowsky21}.  In order to minimize errors in the generation of the random catalog, we only place the random objects within the region of the \textit{HST} and ALMA footprint overlap and ensure the random catalogue has a similar sample size as the real catalogs. To measure the autocorrelation functions, we use the ASTROML function \texttt{bootstrap\_two\_point\_angular} on both the star cluster and GMC catalogs. In order to avoid edge effects, we drop the largest-scale angular bins which correspond to pairs of objects with separations on the order of the size of the field of view. In addition to the full star cluster catalog, we also estimate the autocorrelation functions for the clusters split into two age bins: ${\leq}10$~Myr and ${>}10$~Myr. We then fit power-laws to the autocorrelation functions using a Levenberg--Marquardt non-linear least squares minimization because in a fully hierarchical (or fractal) distribution a smooth power-law decline is expected and indicates a scale-free distribution \citep[][see \S\ref{sec:rd:acf}]{calzetti89}. We test the effects of angular bin sizes by running the analysis with 5, 10, 15, 20, and 25 angular bins and find the same trends for each galaxy; see Figure~\ref{fig:tpcf} for NGC~1566 for a typical example. With fewer bins (i.e., 5 or 10) the main differences are the correlation cannot be measured at small angular scales because the bin size is greater than the angular scale and the autocorrelation functions are lower resolution. With 25 bins, we do not gain any greater resolution at the small angular scales because the bin size is smaller than the actual separation between star cluster pairs. Utilizing 15 or 20 bins appears to provide the optimal spatial resolution for our galaxy sample.  In this figure and the following correlation function figures, the angle $\theta$ is converted to a spatial scale, $r$, in units of parsecs based on the distance to the galaxy.  Analysis of these trends for the full sample is provided in \S\ref{sec:rd:acf}.

\subsection{Angular cross-correlation Functions}
\label{subsec:anl:cross}
For the last step in the analysis, we quantify the excess probability of the clustering of the star clusters with GMCs over a random uniform distribution by estimating the cross-correlation functions following the methodology outlined in Section~\ref{subsec:anl:tpcf}. We generate random catalogs for both the cluster and the cloud populations, and measure the total sample sizes within each catalog:
\begin{align*}
    N_{\rm Dsc} &= \text{Number of real star clusters}\\
    N_{\rm Dgmc} &= \text{Number of real GMCs}\\
    N_{\rm Rsc} &= \text{Number of random catalogue star clusters}\\
    N_{\rm Rgmc} &= \text{Number of random catalogue GMCs}.
\end{align*}
The \citet{landy93} estimator for the cross-correlation is
\begin{equation}\label{eq:landycross1}
    \zeta(\theta) = \frac{ D_{\rm sc}D_{\rm gmc}(\theta) - D_{\rm sc}R_{\rm gmc}(\theta) - R_{\rm sc}D_{\rm gmc}(\theta) + R_{\rm sc}R_{\rm gmc}(\theta) }{R_{\rm sc} R_{\rm gmc}(\theta) }
\end{equation}
where $D_{\rm sc}D_{\rm gmc}(\theta)$ is the number of pairs consisting of one real cluster and one real cloud with a separation~$\theta$, $D_{\rm sc}R_{\rm gmc}(\theta)$ is the number of pairs consisting of one real cluster and one random cloud, $R_{\rm sc}D_{\rm gmc}(\theta)$ is the number of pairs consisting of one random catalogue star cluster and one real GMC, and $R_{\rm sc}R_{\rm gmc}(\theta)$ is the number of pairs consisting of one random catalogue star cluster and one random catalogue GMC. All pair counts are normalized by the total number of objects in the given sample, e.g., $D_{\rm sc}D_{\rm gmc}/N_{\rm Dsc}N_{\rm Dgmc}$. Applying these normalizations to Equation~\eqref{eq:landycross1} and simplifying yields
\begin{equation}\label{eq:landycross2}
\begin{aligned}
    \zeta(\theta) = \left( \frac{N_{\rm Rsc}N_{\rm Rgmc} }{N_{\rm Dsc}N_{\rm Dgmc}} \times \frac{D_{\rm sc}D_{\rm gmc}(\theta) }{R_{\rm sc}R_{\rm gmc}(\theta)} \right) - \\
    \left(\frac{N_{\rm Rsc}}{N_{\rm Dsc}} \times \frac{D_{\rm sc}R_{\rm gmc}(\theta}{R_{\rm sc}R_{\rm gmc}(\theta)} \right) - \left(\frac{N_{\rm Rgmc}}{N_{\rm Dgmc}} \times \frac{R_{\rm sc}D_{\rm gmc}(\theta}{R_{\rm sc}R_{\rm gmc}(\theta)} \right) + 1 .
\end{aligned}
\end{equation}

\begin{figure}
    \includegraphics[width=\columnwidth]{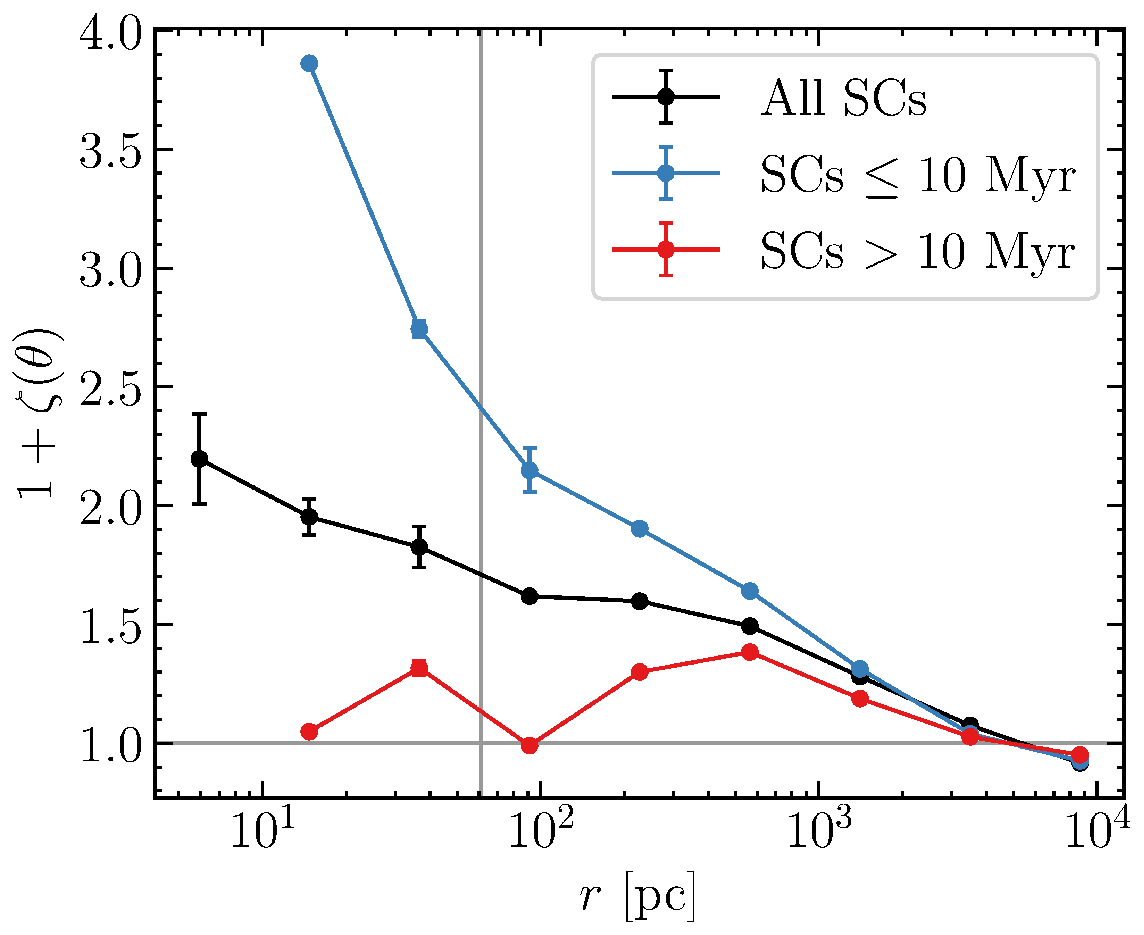}
    \caption{Angular cross-correlation estimate, $1+\zeta(\theta)$, of the star clusters and the GMCs with 10~angular bins for NGC~1566. The function with all star clusters is black, for clusters ${\leq}10$~Myr is blue, and for clusters ${>}10$~Myr is red. The horizontal grey line marks a uniform, random distribution at $1+\zeta(\theta) = 1$. The vertical grey line marks the median radius of the GMCs of NGC~1566. Uncertainties are bootstrap estimates.}
    \label{fig:crosscf}
\end{figure}

Similar to the autocorrelation analysis, we estimate the cross-correlation functions of all the star clusters with the GMCs, clusters ${\leq}10$~Myr with the GMCs, and clusters ${>}10$~Myr with the GMCs. The largest angular bins are dropped in order to minimize edge effects. We also test the cross-correlation estimations using $5, 10, 15, 20, \text{and\ } 25$ angular bins. As with the autocorrelation functions, we find the same general trends for each galaxy regardless of the number of angular bins. Figure~\ref{fig:crosscf} shows the cross-correlation for the star clusters and GMCs of NGC~1566 with 10~angular bins. In contrast to the 20 bins used for the autocorrelation functions, 10 bins was chosen here because there is less data (i.e., fewer star cluster-GMC pairs compared to cluster-cluster pairs).  The higher resolution when using more bins does not provide any additional information.

\section{Results \& Discussion}
\label{sec:rd}

\subsection{Nearest Neighbour Analysis}
\label{subsec:rd:nn}

\begin{figure*}
    \centering
    \includegraphics[width=\textwidth]{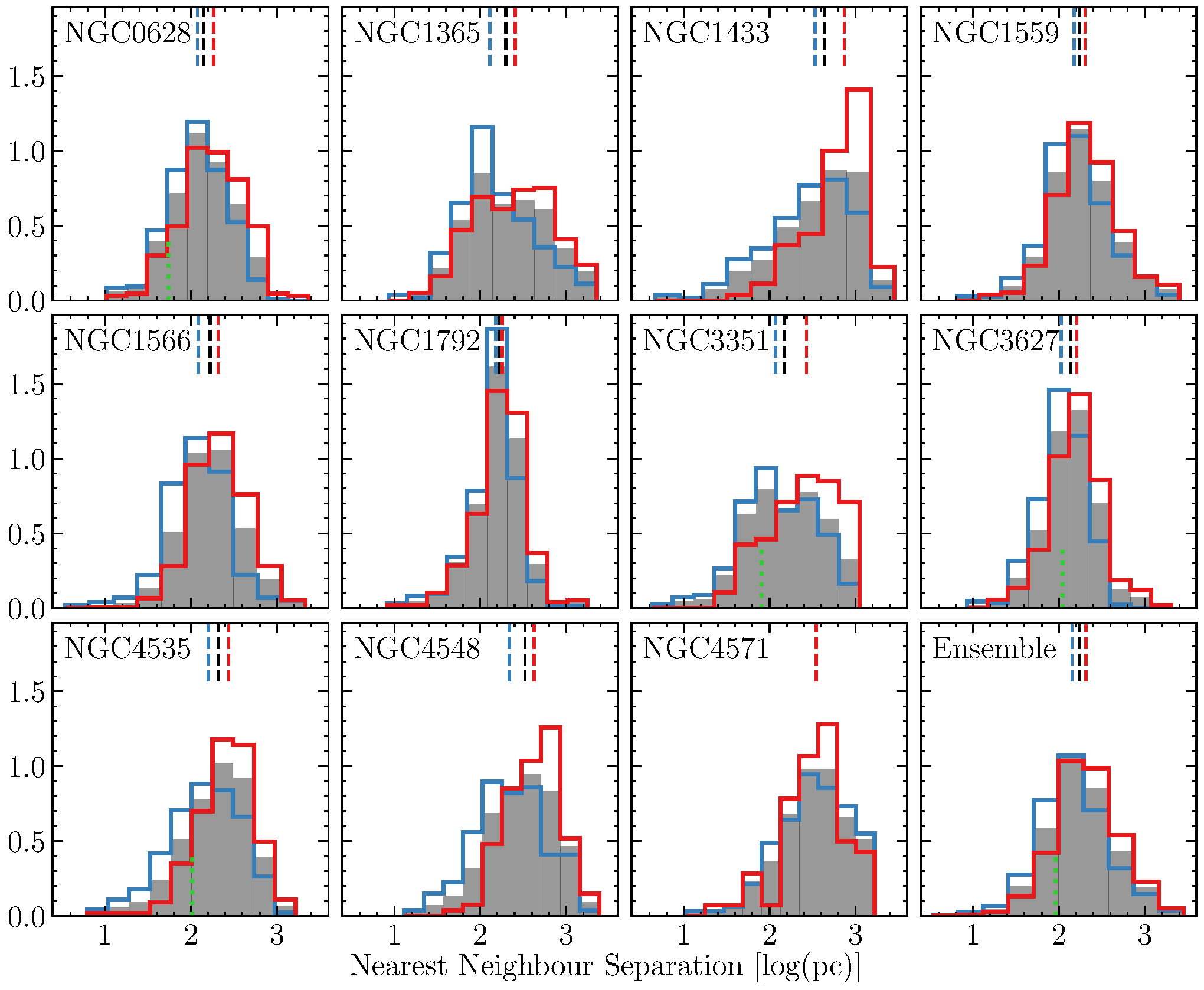}
    \caption{Histograms of the separations between star clusters and the centre of the nearest neighbour GMC for each of the 11~galaxies and an ensemble of all the galaxies. The distribution for the full sample of star clusters is shown in grey, for clusters ${\leq}10$~Myr is blue, and for clusters ${>}10$~Myr is red. Median separations for each distribution are marked with dashed lines.  Theoretical estimates for the nearest neighbor separations for NGC~0628, NGC~3351, NGC~3627, NGC~4535, and the ensemble are provided by green dotted lines (see text).  The histograms are normalized by the total number of counts and the bin width for each histogram.}
    \label{fig:1nn_grid}
\end{figure*}

Figure~\ref{fig:1nn_grid} shows the histograms of the separations between the star clusters and the nearest neighbour GMC for all 11~galaxies. Included for each galaxy is the histogram for all star clusters and for the clusters split into the two age bins: ${\leq}10$~Myr (blue) and ${>}10$~Myr (red). For all star clusters, the median separations $\tilde{\Delta r}$ to the nearest neighbour GMC range from ${\sim}1.8\arcsec$ (NGC~1559) to $4.9\arcsec$ (NGC~4571). In all galaxies except NGC~4571, the young clusters are found to be closer to their nearest neighbour GMCs than the older populations.  The ``signal-to-noise'' in these median separations can be estimated via
\begin{equation}
S/N (\tilde{\Delta r}) = \frac {\Tilde{\Delta r}_{\rm old} - \tilde{\Delta r}_{\rm young}} {\sqrt{\sigma^2_{\rm old} + \sigma^2_{\rm young}}}
\end{equation}
where $\sigma_{\rm old}$ and $\sigma_{\rm young}$ represent the standard deviations in the median values for 1000 bootstrapped samples for the older and younger age bins, respectively.  Except for NGC~4571, the $S/N$ values lie in the range 4--8.  In NGC~4571, the median separations for the three histograms are nearly the same. This result is due to the flocculent nature of NGC~4571---unlike the other 10 galaxies in this sample, where the CO emission tends to align with the spiral arm structures that dominate the optical morphologies, the GMCs and stellar clusters in NGC~4571 exhibit much patchier and more spatially uniform distributions. NGC~1433 and NGC~3351 have the greatest differences in separations between the young and old populations. Both of these galaxies show evidence of recent star formation at their centres with a strong CO concentration there.  

Finally, we include in certain sub-panels of Figure~\ref{fig:1nn_grid} theoretical estimates for the expected nearest neighbour separations $r_n$.  Broadly following the approach described in \citet{kruijssen19} and the parameter values presented in \citep{chevance20}, the expected separations are $r_n\approx 0.443 \lambda \sqrt{\tau/t_{\rm gas}}$ where $\lambda$ is the typical separation between independent star-forming regions, $\tau$ is the evolutionary timescale for star formation overall, and $t_{\rm gas}$ is the timescale for the molecular clouds in particular.  The correction factor $\sqrt{\tau/t_{\rm gas}}$ ($\approx1.1$) accounts for the fact that we specifically require the neighbour to be a GMC, which only covers part of the region timeline described in \citet{chevance20}.  The medians for the distributions presented in Figure~\ref{fig:1nn_grid} are somewhat larger than the theoretical expectations, presumably because the GMC catalogues have lower effective spatial resolution than the maps themselves.

We check the second and third nearest neighbour separation histograms and find the same trends where the young stellar populations lie closer to the GMCs than the older clusters. 

\subsection{Star Clusters Associated with GMCs}
\label{subsec:rd:assoc}

\begin{figure*}
    \centering
    \includegraphics[width=\textwidth]{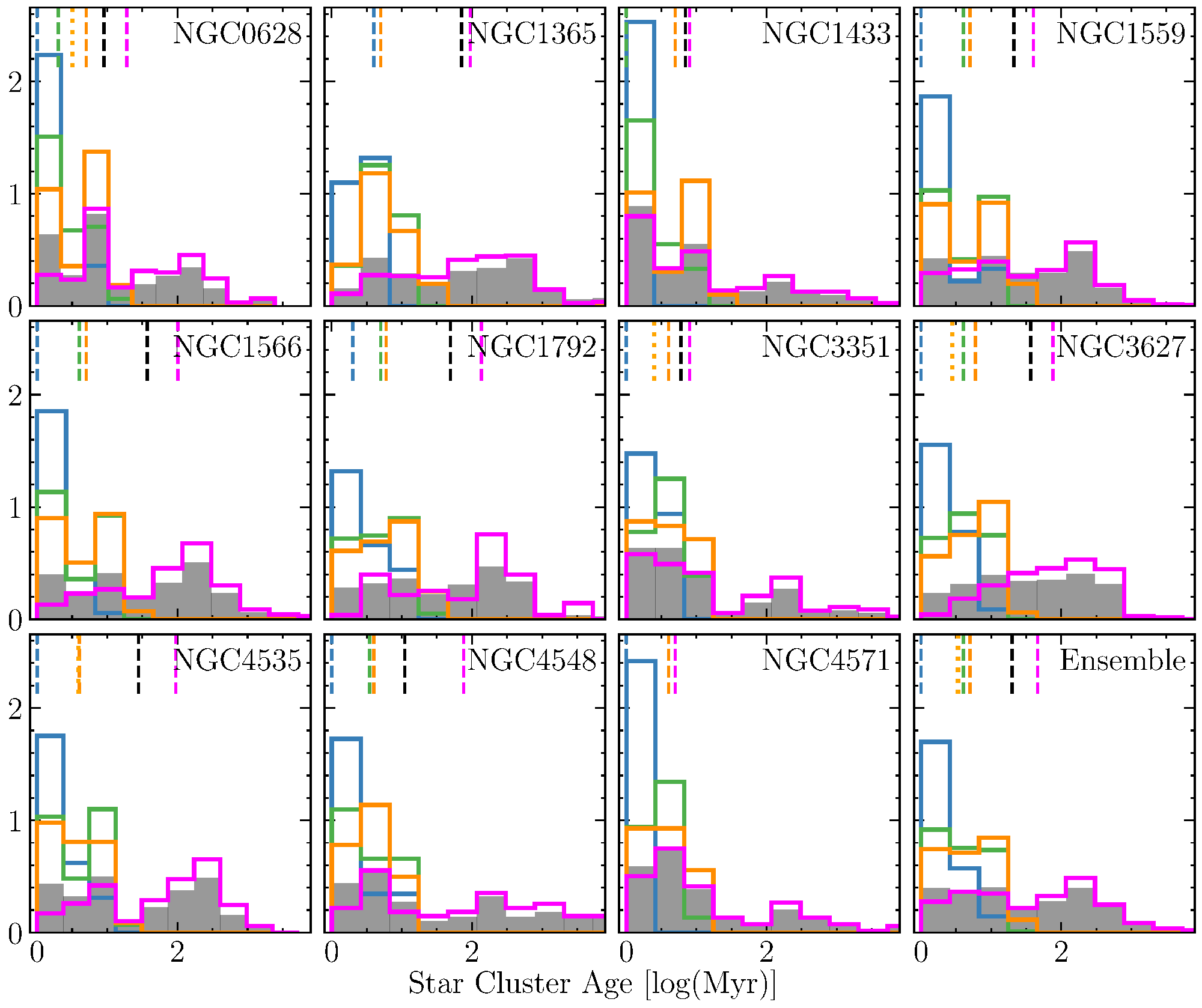}
    \caption{Histograms of the star cluster ages split on how the star cluster is associated with the closest GMC: within 1\,$R_{\mathrm{GMC}}$ (blue), between 1 and 2\,$R_{\mathrm{GMC}}$ (green), between 2 and 3\,$R_{\mathrm{GMC}}$ (orange), or beyond 3\,$R_{\mathrm{GMC}}$ which is considered unassociated (fuchsia). The distribution of the ages for all star clusters is in grey. Median ages for each distribution are marked with dashed lines. An assumed velocity cutoff of $10$~km~s$^{-1}$ is applied to minimize the chance alignment of clusters with GMCs if the clusters have had enough time to cross into the line-of-sight.  The histograms are normalized by the total number of counts and the bin width for each histogram.  For reference we include for specific targets plus the ensemble the estimated feedback timescales \citep{chevance20} for dispersing the natal gas after stellar clusters first become visible (vertical dotted orange lines).}
    \label{fig:assoc_grid}
\end{figure*}

In Section~\ref{subsec:anl:sc_gmc}, we describe our method for associating the star clusters with GMCs while taking into account the sizes of the GMCs. Figure~\ref{fig:assoc_grid} shows the distribution of the star cluster ages split by how the clusters are associated with the closest GMC and Table~\ref{tab:gals_assoc} shows the median cluster ages.\footnote{We have incorporated a $10$~km~s$^{-1}$ velocity cutoff in our analysis to account for clusters that may have drifted during their lifetime into a chance line-of-sight alignment with a non-natal GMC; we require all cluster-GMC associations to satisfy $v \times t({\rm age}) < 3 R_{\rm GMC}$.  We acknowledge, though, that this a coarse approximation and is limited by the fact that GMC radii are highly related to the resolution of the CO data, and the spatial resolution of the CO data affects the minimum peak spacing.  Moreover, the GMCs crowd on few hundred parsec scales and given their typically sizes, there may be multiple associations for a given cluster. In these cases, we opt for the most massive GMC to be the associated one. These are the issues that make this inherently complicated.}  Across all galaxies, the median cluster ages are consistently the youngest when within the radius of the closest GMC. As the clusters increase in distance from the closest GMC, the median ages increase. For clusters between 2 and 3 GMC radii, the median ages are no greater than 6~Myr across all galaxies (the range is $4-6$~Myr, with uncertainties for a given galaxy on the order of 0.5~Myr; see Table~\ref{tab:gals_assoc}).  Given these results, after about 6~Myr, the star clusters have had enough time to no longer be associated with their natal GMCs. This result is consistent with the measurement from \citet{grasha19} in M51 using the LEGUS star clusters and \mbox{CO(1--0)} maps from PAWS.  Similar timescales were found by \cite{kawamura09} for the Large Magellanic Cloud and \cite{corbelli17} for M~33 for the duration of GMCs being associated with non-embedded stellar clusters. 

\newcommand\mcone[1]{\multicolumn{1}{c}{#1}}
\newcommand\mctwo[1]{\multicolumn{2}{c}{#1}}
\begin{table*}
\sisetup{separate-uncertainty,zero-decimal-to-integer}
\centerline{Table 2. Median Star Cluster Ages}
\begin{tabular}{l S@{$\, \,$}l S@{$\, \,$}l S@{$\, \,$}l S@{$\, \,$}l S@{$\, \,$}l }
\hline\hline
\mcone{Galaxy} & \mctwo{All SCs} & \mctwo{$0 < r/R_{\text{GMC}} \leq 1$} & \multicolumn{2}{r}{$1 < r/R_{\text{GMC}} \leq 2$} & \multicolumn{2}{r}{$2 < r/R_{\text{GMC}} \leq 3$} & \mctwo{Unassociated} \\
\hline
NGC~0628 & 09 \pm 0.32 & (684) & 01 \pm 0.33 & (33) & 02 \pm 0.32 & (092) & 05 \pm 0.44 & (142) & 018 \pm 2.54 & (266) \\
NGC~1365 & 70 \pm 4.96 & (633) & 04 \pm 0.92 & (11) & 05 \pm 0.38 & (054) & 05 \pm 0.54 & (098) & 095 \pm 7.23 & (264) \\
NGC~1433 & 07 \pm 0.62 & (293) & 01 \pm 0.60 & (09) & 01 \pm 0.56 & (023) & 05 \pm 1.39 & (025) & 007 \pm 0.67 & (219) \\
NGC~1559 & 21 \pm 1.24 & (926) & 01 \pm 0.25 & (22) & 04 \pm 0.56 & (087) & 05 \pm 0.54 & (160) & 040 \pm 2.43 & (387) \\
NGC~1566 & 37 \pm 2.31 & (838) & 01 \pm 0.25 & (43) & 04 \pm 0.61 & (081) & 05 \pm 0.60 & (134) & 101 \pm 6.28 & (335) \\
NGC~1792 & 48 \pm 3.25 & (669) & 02 \pm 0.63 & (22) & 05 \pm 0.58 & (094) & 06 \pm 0.46 & (147) & 135 \pm 23.8 & (067) \\
NGC~3351 & 06 \pm 0.41 & (396) & 01 \pm 0.65 & (18) & 04 \pm 0.45 & (056) & 04 \pm 0.62 & (061) & 007 \pm 0.39 & (221) \\
NGC~3627 & 37 \pm 1.46 & (948) & 01 \pm 0.42 & (28) & 04 \pm 0.43 & (100) & 06 \pm 0.53 & (164) & 076 \pm 3.71 & (287) \\
NGC~4535 & 28 \pm 3.25 & (411) & 01 \pm 0.51 & (26) & 04 \pm 0.85 & (039) & 04 \pm 0.60 & (063) & 095 \pm 8.41 & (185) \\
NGC~4548 & 10 \pm 3.49 & (237) & 01 \pm 0.76 & (14) & 03 \pm 0.74 & (022) & 04 \pm 0.48 & (034) & 076 \pm 13.1 & (131) \\
NGC~4571 & 05 \pm 0.29 & (213) & 01 \pm 0.90 & (06) & 04 \pm 0.79 & (018) & 04 \pm 0.88 & (013) & 005 \pm 0.54 & (164) \\
\hline
Ensemble & 19 \pm 0.61 & (6248) & 01 \pm 0.18 & (232) & 04 \pm 0.17 & (666) & 05 \pm 0.18 & (1041) & 046 \pm 1.53 & (2526) \\
\end{tabular}
\caption{All cluster ages given in Myr. Uncertainties on the medians are bootstrap estimates based on the uncertainties of the cluster ages. The number of clusters in each sample is given in parentheses.}
\label{tab:gals_assoc}
\end{table*}

This timescale is related to the feedback timescale $t_{\mathrm{fb}}$---the time it takes for feedback mechanisms to disperse the natal gas after stellar clusters first become visible (unembedded). This feedback timescale is not a measure of the GMC lifetime, which is found to be short, on the order of 10~Myr to 30~Myr, but the time it takes for star clusters to no longer be associated with their natal GMCs. Given the short lifetimes of GMCs, finding younger star clusters closer to their natal GMCs implies that the stellar hierarchical distribution is inherited from the GMCs and the interstellar medium. Our measurement tracks well with the measurements from the ``uncertainty principle of star formation'' for NGC~0300 \citep{kruijssen19} and the samples of nearby galaxies studied by \cite{chevance20} and \cite{kim21} (four of the nine galaxies analyzed by \cite{chevance20} are also in our sample). As described above, for our ensemble of 11 galaxies the clusters are typically on the cusp of no longer being associated with GMCs after $\sim$6~Myr.  By comparison, \cite{kim21} find ${\sim}3$~Myr for the typical embedded phase (see also \citealt{benincasa20}) and \citet{chevance20} and \citet{kim22} find $t_{\mathrm{fb}} \approx 3$~Myr for the unembedded phase, for a total of ${\sim}7$~Myr spanning the embedded and unembedded phases; accounting for uncertainties and intrinsic variation between galaxies, our inferred ensemble-wide timescale for clusters is consistent with the sum of the embedded and unembedded timescales found by \citet{chevance20}, \citet{kim21}, and \citet{kim22}.

For the clusters lying beyond three GMC radii from the GMC centroids, the median ages range greatly across the 11~galaxies from 5~Myr (NGC~4571) to 136~Myr (NGC~1792).  Interestingly, the galaxy in our sample with the sparsest (lowest surface density) CO distribution, NGC~4571, has unassociated stellar clusters that skew youngest compared to the other galaxies. Similarly, in NGC~1433 and NGC~3351 where the CO is concentrated in the centres, the median unassociated cluster age is younger (7~Myr) than the rest of the galaxies. In all galaxies, the median unassociated cluster age is still greater than the median age of all clusters.  However, we must acknowledge that this trend is influenced by our implementation of a velocity cutoff.

\subsection{Dependence on Galactic Environment}
    \label{subsec:rd:env}

\begin{figure*}
    \centering
    \includegraphics[width=\textwidth]{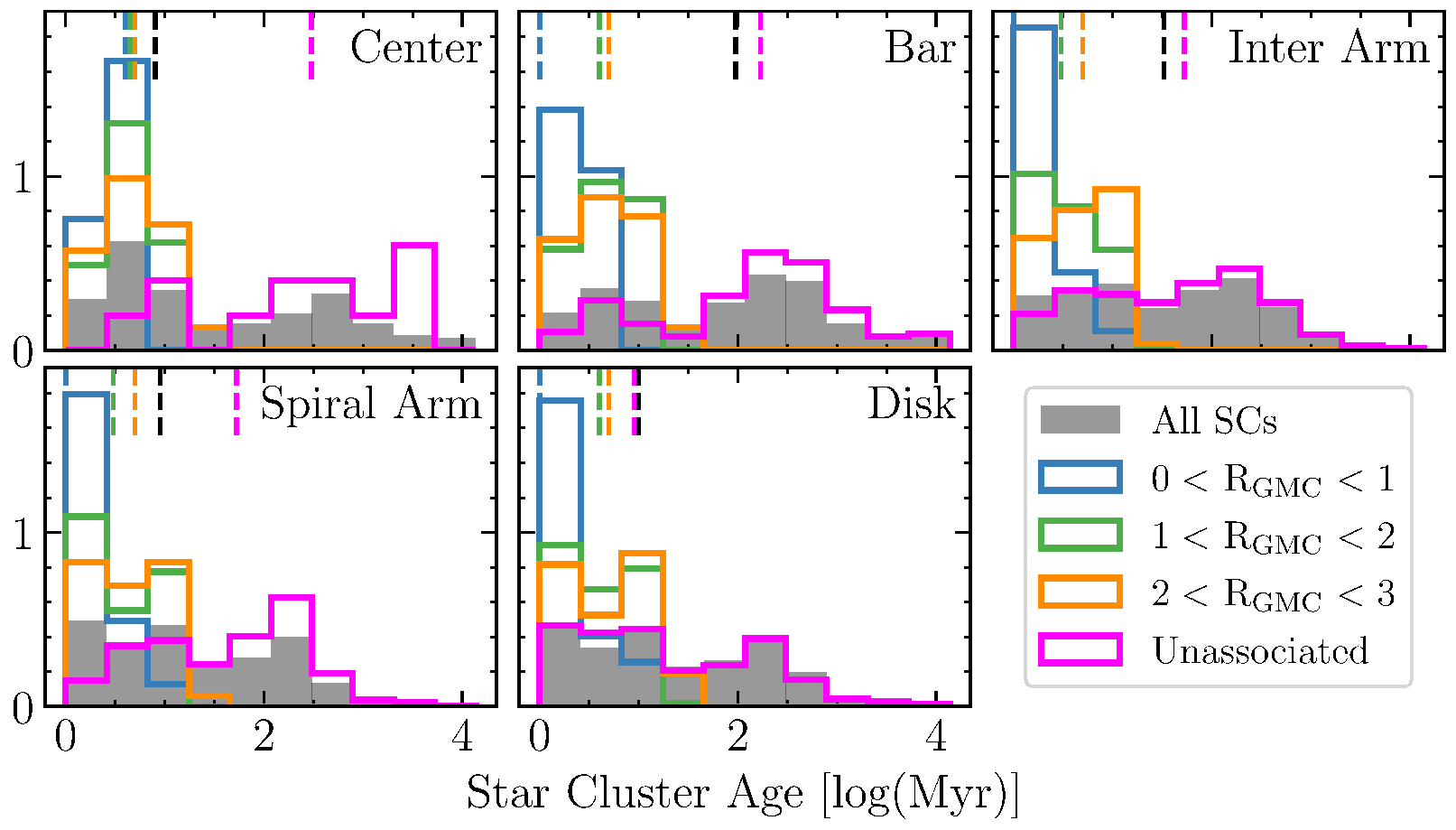}
    \caption{Same as Figure~\ref{fig:assoc_grid} except star clusters are grouped by their location with the galaxies: galaxy centre, bars, interarm, spiral arms, and disk. Median cluster ages are marked with dashed lines. A $10$~km~s$^{-1}$ velocity cutoff is applied. }
    \label{fig:assoc_env_grid}
\end{figure*}

\begin{table*}
\sisetup{separate-uncertainty,zero-decimal-to-integer}
\centerline{Table 3. Median Star Cluster Ages by Galactic Environment}
\begin{tabular}{l S@{$\, \,$}l S@{$\, \,$}l S@{$\, \,$}l S@{$\, \,$}l S@{$\, \,$}l }
\hline\hline
\mcone{Environment} & \mctwo{All SCs} & \mctwo{$0 < r/R_{\text{GMC}} \leq 1$} & \multicolumn{2}{r}{$1 < r/R_{\text{GMC}} \leq 2$} & \multicolumn{2}{r}{$2 < r/R_{\text{GMC}} \leq 3$} & \mctwo{Unassociated} \\
\hline
Center     & 07 \pm 2.59 & (0377) & 04 \pm 0.67 & (16) & 04 \pm 0.26 & (074) & 05 \pm 0.44 & (110) & 301 \pm 138.6 & (012) \\
Bar        & 95 \pm 5.01 & (0693) & 01 \pm 0.93 & (14) & 04 \pm 0.52 & (500) & 05 \pm 0.51 & (110) & 168 \pm 10.7 & (301) \\
Inter Arm  & 32 \pm 1.37 & (1654) & 01 \pm 0.29 & (43) & 03 \pm 0.34 & (117) & 05 \pm 0.41 & (180) & 053 \pm 2.51 & (969) \\
Spiral Arm & 09 \pm 0.30 & (1420) & 01 \pm 0.21 & (93) & 03 \pm 0.22 & (206) & 05 \pm 0.32 & (306) & 053 \pm 3.43 & (388) \\
Disk       & 10 \pm 0.90 & (2077) & 01 \pm 0.25 & (66) & 04 \pm 0.29 & (219) & 05 \pm 0.36 & (335) & 009 \pm 0.28 & (829) \\
\hline
\end{tabular}
\caption{All cluster ages given in Myr. Uncertainties on the medians are bootstrap estimates based on the uncertainties of the cluster ages. The number of clusters in each sample is given in parentheses.}
\label{tab:env_assoc}
\end{table*}

We can study how the galactic environment of the star clusters and GMCs affect the timescales of star formation by using the environmental masks developed by the PHANGS collaboration. Details on how the masks are produced are given in \citet{querejeta21}. In short, discs and bulges are identified using $3.6$~$\mu$m images from the \textit{Spitzer} Survey of the Stellar Structure in Galaxies (S$^{4}$G) pipeline or ancillary \textit{Spitzer} images following \citet{salo15} and \citet{herrera-endoqui15}. Near-infrared images are used to identify bars and rings, and spiral arms are defined by fitting a log-spiral function to bright regions along the arms. Arm widths are defined by the CO emission measured in PHANGS--ALMA maps. For this study, we choose to use the masks which separate the galactic environments into the centre, bar, interarm, spiral arms, and disk. The masks do provide further differentiation of environments for example, bars and bar ends, but we opt for the more simplified masks. 

Figure~\ref{fig:assoc_env_grid} shows the distribution of star cluster ages across the five environmental masks and Table~\ref{tab:env_assoc} gives the median ages with uncertainties. Across all five environments, the median cluster age is 5~Myr for clusters between 2 and 3~GMC radii. This is consistent with the time scale for a cluster to disassociate from its natal gas cloud as found in Section~\ref{subsec:rd:assoc}.

Galaxies with a centre region identified are NGC~0628, NGC~1365, NGC~1433, NGC~1792, NGC~3351, NGC~3627, NGC~4535, NGC~4548, NGC~4571. The clusters which lie in the centres are found to be young (${<}10$~Myr) and the trend of cluster age increasing with distance from closest GMC is seen. There are only 12 unassociated clusters which have a median age of $300$~Myr; this is significantly older than the unassociated population of the other environments and any single galaxy. These old unassociated clusters are likely old globular clusters which reside in the stellar bulges. For the star clusters of NGC~3351, \citet{turner21} identify globular cluster candidates in the bulge and show the SED fitting return underestimated ages for the globular clusters. Ages around 10~Gyr are expected while the SED fitting gives ages of a few 100~Myr. This is likely why the clusters identified here are as old as expected if they are indeed globular clusters.

For clusters in the bars of the barred galaxies (NGC~1365, NGC~1433, NGC~1559, NGC~1566, NGC~3351, NGC~3627, NGC~4535, NGC~4548), the median age is relatively old at nearly 100~Myr. Clusters associated with GMCs are still found to be the youngest. Similar to the centre clusters, the unassociated population is old with median age of $169$~Myr. The bar clusters trending old is mostly likely due to a lack of recent star formation as bars act to funnel gas into the centres of the galaxies which triggers star formation there.

The environmental masks show that all galaxies except NGC~4571 have spiral arms or, in the case of NGC~3351, a ring. For clusters in both the interarm regions and spiral arms, we see the same trend with the youngest clusters (${\sim}1$~Myr) associated closely with GMCs and trending slightly older as they move away from their natal gas clouds. Unassociated clusters in both regions give a median age around 50~Myr. However, spiral arm clusters are found to be younger on average than the interarm clusters due to more recent star formation occurring directly within the spiral arms. The difference in the median ages of spiral and interarm clusters suggests it takes ${\sim}20$~Myr for clusters to migrate out of the spiral arms.

Finally, clusters in the outer disks of the galaxies are found to be mostly young with a median age of $10$~Myr. The same trend of cluster age increasing with distance to associated GMC is seen again. Interestingly, the unassociated clusters in the discs have a very similar age distribution to the full sample of disc clusters and a median age just under $10$~Myr. This distribution is likely skewed young because of NGC~4571 where essentially all clusters are considered to be in the disc and the relative lack of CO means most clusters are unassociated with a GMC. 

\subsubsection{Dependence on Spiral Arm Structure}
\label{sec:rd:spar}
As shown Figure~\ref{fig:assoc_env_grid}, clusters residing in spiral arms are found to be very young and closely associated with GMCs. We check for dependence of this trend on the spiral arm structure, specifically on the pitch angle of the spirals. Generally, galactic spiral arms are well approximated by logarithmic spirals \citep[][]{kennicutt81}. The logarithmic spirals are of the form 
\begin{equation}\label{eq:spiral}
    R = R_{0} e^{\theta \tan(\phi)}
\end{equation}
where $R$ is the radius of the spiral, $\theta$ is the azimuthal angle, $R_{0}$ is the initial radius at $\theta = 0$, and $\phi$ is the pitch angle. Equation~\ref{eq:spiral} can be linearized to the form
\begin{equation}\label{eq:linspiral}
    \ln(R) = \ln(R_{0}) + \theta\tan(\phi).
\end{equation}
The environmental masks in \citet{querejeta21} provide the slope, $\theta$, and intercept, $\ln(R_{0})$, of best-fit logarithmic spirals for each of the identified spiral arms which we then used to obtain the spiral arm pitch angles. There are six galaxies with identified spiral arms--NGC~0628 (6 arms), NGC~1365 (1 arm), NGC~1566 (4 arms), NGC~3627 (4 arms), NGC~4535 (3 arms), NGC~4548 (5 arms).

\begin{figure}
    \centering
    \includegraphics[width=\columnwidth]{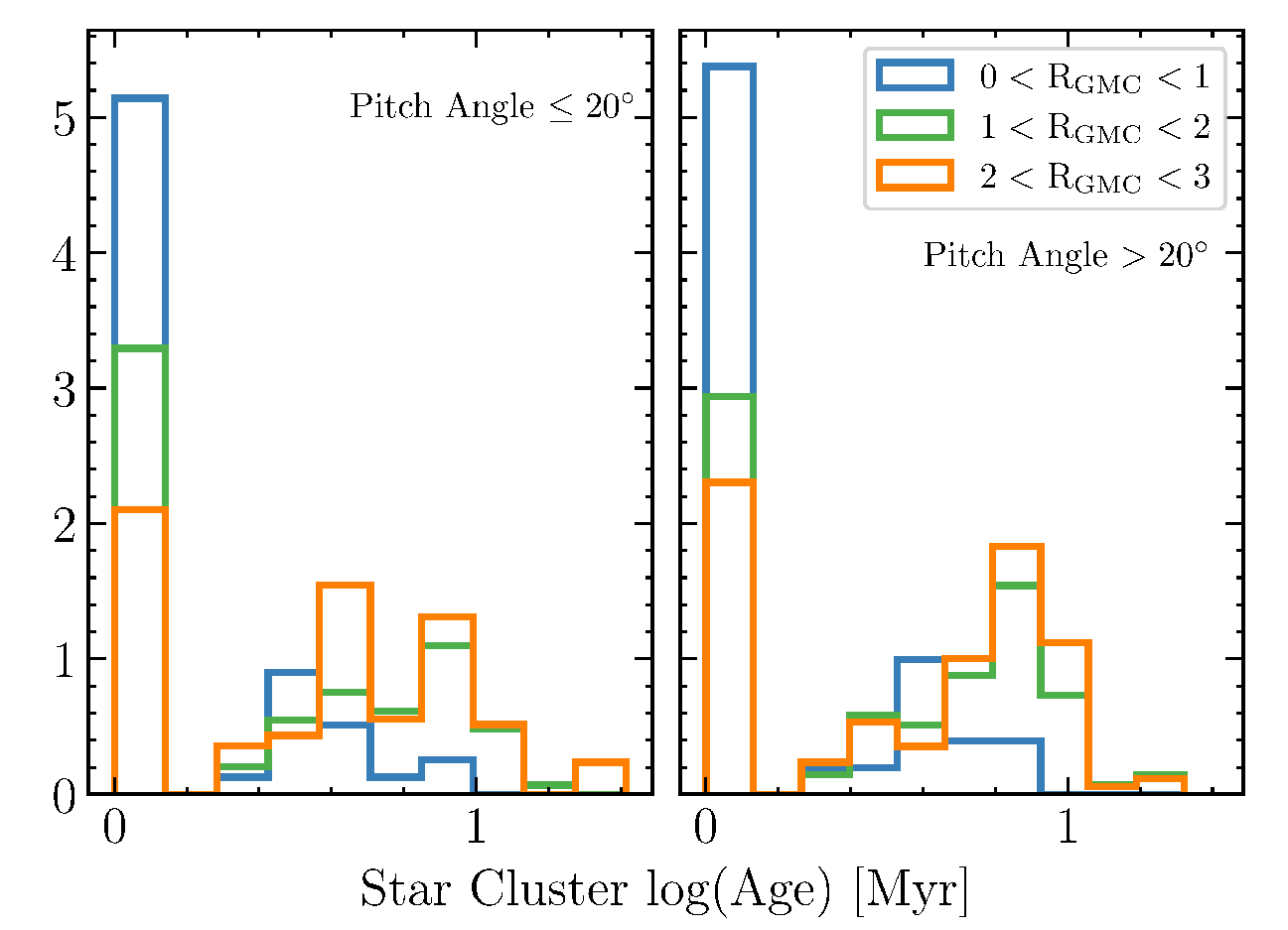}
    \caption{Histograms of star cluster ages for all clusters which lie within a spiral arm with pitch angle less than or equal to 20 degrees (left) and greater than 20 degrees (right). Histograms are color-coded by how the star cluster is associated with the nearest GMC following the same scheme as Figures~\ref{fig:assoc_hist},~\ref{fig:assoc_grid}, and~\ref{fig:assoc_env_grid}.}
    \label{fig:pitchangle}
\end{figure}

Across the galaxy sample, we check the star cluster age and star cluster--GMC separation for all clusters which lie within a spiral arm; results are shown in Figure~\ref{fig:pitchangle}. Spiral arms are split by pitch angle at 20 degrees. We find that the star cluster age histograms are essentially no different between spiral arms with tighter and looser pitch angles. A division at 30~degrees was also checked and showed no dependence on the pitch angle dividing threshold. We limit this analysis to only clusters found within spiral arms as the pitch angle of spiral arms within individual galaxies can range significantly. For example, the six spiral arms measured in NGC~0628 have pitch angles ranging from 11 degrees to 30 degrees. This greatly confuses the analysis if it is to be applied to star clusters outside of the spiral arms. Given these results, we find no dependence of star cluster age and star cluster--GMC separation on the spiral arm pitch angle.

\subsection{Angular Two-Point Correlation Functions}
\label{sec:rd:acf}

\begin{figure*}
    \centering
    \includegraphics[width=\textwidth]{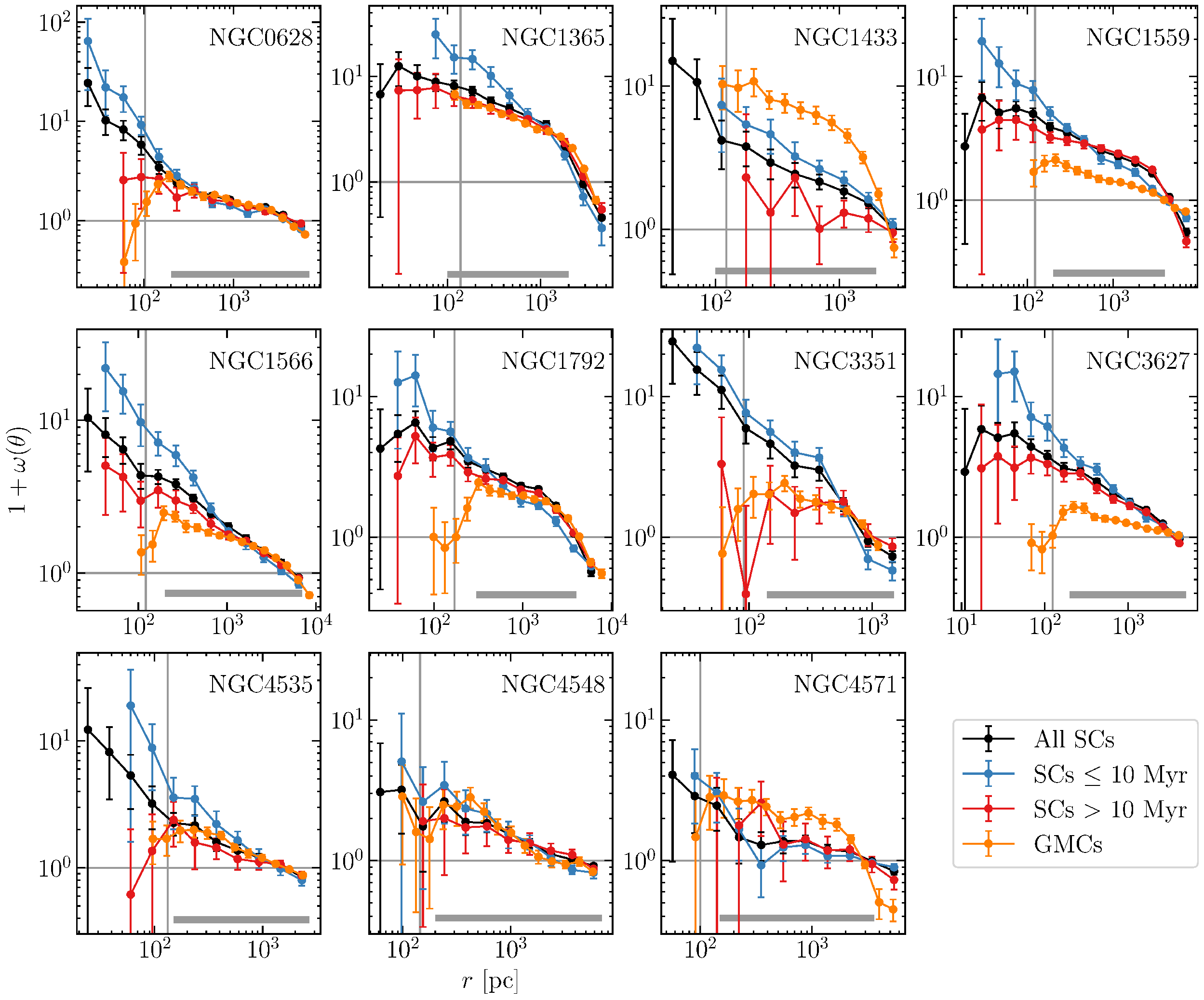}
    \caption{Two-point (auto) correlation functions for all star clusters (black), clusters ${\leq}10$~Myr (blue), clusters ${>}10$~Myr (red), and GMCs (orange). The line at $1 + \omega(\theta) = 1$ marks where a uniform, random distribution. Twice the median radius of the GMCs in each galaxy is shown as the vertical grey line. Uncertainties are bootstrap estimates. The thick grey line marks the distance range over which the power-laws are fit; the best-fitting power-law slopes are given in Table~\ref{tab:tpcf_plaw}. Correlation functions which exhibit anti-correlation ($1 + \omega(\theta) < 1$) are most likely artifacts of the random sampling i.e., there are more random points than data points for that angular bin. In which case, correlation at these bins can be considered consistent with a random distribution.} 
    \label{fig:tpcf_grid}
\end{figure*}

\begin{table*}
\sisetup{separate-uncertainty,table-auto-round=true,round-mode=places,round-precision=2}
\centerline{Table 4. Two-Point Autocorrelation Functions: Best-Fitting Power Law Slopes and Amplitudes at 300~pc}
\begin{tabular}{ccccccccc}
\hline\hline
\mcone{Galaxy} & {All SCs} & {SCs $\leq10$ Myr} & {SCs $>10$ Myr} & {GMCs}  & {All SCs} & {SCs $\leq10$ Myr} & {SCs $>10$ Myr} & {GMCs}\\
\mcone{} & {$\alpha$} & {$\alpha$} & {$\alpha$} & {$\alpha$} & {$1+\omega(\theta)$} & {$1+\omega(\theta)$} & {$1+\omega(\theta)$} & {$1+\omega(\theta)$} \\
\hline
NGC~0628 & $-$0.28$\pm$0.02 & $-$0.39$\pm$0.05 & $-$0.20$\pm$0.03 & $-$0.24$\pm$0.04 & 2.20$\pm$0.20 & 2.46$\pm$0.35 & 1.86$\pm$0.37 & 2.15$\pm$0.19\\
NGC~1365 & $-$0.41$\pm$0.03 & $-$0.66$\pm$0.09 & $-$0.33$\pm$0.02 & $-$0.47$\pm$0.11 & 5.75$\pm$0.53 & 9.98$\pm$2.10 & 5.04$\pm$0.63 & 4.97$\pm$0.28 \\
NGC~1433 & $-$0.38$\pm$0.02 & $-$0.55$\pm$0.02 & $-$0.28$\pm$0.17 & $-$0.82$\pm$0.14 & 2.86$\pm$0.66 & 4.43$\pm$1.19 & 1.44$\pm$1.48 & 7.96$\pm$1.15 \\
NGC~1559 & $-$0.32$\pm$0.02 & $-$0.49$\pm$0.03 & $-$0.22$\pm$0.02 & $-$0.22$\pm$0.02 & 3.51$\pm$0.26 & 3.77$\pm$0.34 & 3.05$\pm$0.26 & 1.88$\pm$0.18 \\
NGC~1566 & $-$0.44$\pm$0.02 & $-$0.73$\pm$0.05 & $-$0.36$\pm$0.02 & $-$0.26$\pm$0.03 & 3.64$\pm$0.28 & 5.52$\pm$0.81 & 2.92$\pm$0.42 & 2.19$\pm$0.18 \\
NGC~1792 & $-$0.36$\pm$0.06 & $-$0.51$\pm$0.05 & $-$0.32$\pm$0.06 & $-$0.49$\pm$0.06 & 3.29$\pm$0.28 & 3.43$\pm$0.58 & 2.78$\pm$0.33 & 2.27$\pm$0.29 \\
NGC~3351 & $-$0.73$\pm$0.10 & $-$0.84$\pm$0.15 & $-$0.29$\pm$0.11 & $-$0.76$\pm$0.02 & 3.12$\pm$0.53 & 3.85$\pm$0.70 & 1.61$\pm$0.65 & 1.83$\pm$0.23 \\
NGC~3627 & $-$0.38$\pm$0.01 & $-$0.47$\pm$0.03 & $-$0.38$\pm$0.02 & $-$0.08$\pm$0.02 & 2.84$\pm$0.16 & 3.28$\pm$0.36 & 2.73$\pm$0.25 & 1.60$\pm$0.12 \\
NGC~4535 & $-$0.41$\pm$0.04 & $-$0.73$\pm$0.06 & $-$0.25$\pm$0.03 & $-$0.23$\pm$0.02 & 1.90$\pm$0.35 & 2.89$\pm$0.76 & 1.52$\pm$0.45 & 1.90$\pm$0.20 \\
NGC~4548 & $-$0.34$\pm$0.03 & $-$0.50$\pm$0.04 & $-$0.24$\pm$0.02 & $-$0.16$\pm$0.04 & 2.33$\pm$0.60 & 2.99$\pm$1.28 & 1.89$\pm$0.96 & 2.44$\pm$0.69 \\
NGC~4571 & $-$0.11$\pm$0.03 & $-$0.14$\pm$0.08 & $-$0.28$\pm$0.11 & $-$0.59$\pm$0.11 & 1.36$\pm$0.39 & 1.23$\pm$0.50 & 2.26$\pm$1.25 & 2.67$\pm$0.48 \\
\hline
\end{tabular}
\caption{A Levenberg--Marquardt non-linear least squares minimization is used to fit a power-law of the form $A\theta^{\alpha}$ to the autocorrelation functions shown in Figure~\ref{fig:tpcf_grid}. The power-laws are constrained to the distance ranges marked by the thick grey lines in Figure~\ref{fig:tpcf_grid}.}
\label{tab:tpcf_plaw}
\end{table*}

We quantify how clustered the spatial distributions of the star clusters and GMCs are using two-point autocorrelation functions as detailed in Section~\ref{subsec:anl:tpcf}. Figure~\ref{fig:tpcf_grid} shows the autocorrelation functions for each galaxy broken down as: all star clusters, clusters ${\leq}10$~Myr, clusters ${>}10$~Myr, and all GMCs\footnote{No significant differences are found if the age threshold is lowered to 5~Myr.}. In all galaxies except for NGC~4571, the young clusters are found to be more highly correlated over small spatial scales than the older populations. At larger scales on the order of several kiloparsecs, the correlation functions for both the young and old cluster populations become essentially equal. This means the correlation lengths, the scale at which the distribution is random ($1+\omega(\theta) = 1)$, are equal for both populations. Given these results, the young populations still show fractal nature of hierarchical star formation which dissolves with time as shown by the autocorrelation functions of the older cluster populations.

Table~\ref{tab:tpcf_plaw} gives the best-fitting power-law slopes for each of the measured autocorrelation functions. For the total star cluster populations across all galaxies, reasonably smooth power-law fits are found which demonstrates the scale-free, hierarchical (or fractal) structure of the stellar distribution. Except for the flocculent galaxy NGC~4571, each galaxy shows a steeper power-law slope for the younger clusters than for the older clusters.  The galaxies with the largest difference in slope between the young and older populations, for the selected spatial ranges for the fits, are NGC~1566, NGC~3351, and NGC~4535. These three galaxies all possess very distinct spiral arms and rings where the GMCs and younger clusters are mostly concentrated.

Since the higher correlation of the young clusters is thought to be a result of inheriting the hierarchical structure of the interstellar medium, we expect the GMCs to show equally high correlation. However, in most cases, the distribution of the GMCs is found to be much less correlated; the GMCs seem to be closer to randomly distributed. \citet{grasha19} note a similar result for the GMCs observed in M51. This suggests either (1) there are GMCs which have not yet formed star clusters and are therefore dampening the correlation signal; or (2) the ultraviolet-optical HST data are insensitive to clusters less massive than ${\sim}10^3 {-} 10^4~M_\odot$ (see \S\ref{sec:data}) or to heavily buried stellar clusters (a follow-up project with the James Webb Space Telescope has been approved for 19 PHANGS targets, to uncover and characterize such embedded systems).

In a number of galaxies, the GMC correlation is roughly random at the smallest scale, then increases with scale length until a peak at intermediate separations, then declines again. This peak in the correlation most likely results from CPROPS combining overlapping GMCs into a single GMC (if they have similar velocities), which prevents us from finding a significant number of GMCs with separations less than twice the typical GMC radius.  The value of $2 R_{\mathrm{GMC}}$ is presented in each panel of Figure~\ref{fig:tpcf_grid}, where we would expect to start noticing the impact of the cloud identification on the autocorrelation function (and thus explains the downturn in the GMC autocorrelation function at small scales in most galaxies).  In NGC~1433, the GMCs are concentrated at the centre of the galaxy and at the end of the bars. This leads to high correlation at small scales. Although NGC~3351 has a similar CO distribution as NGC~1433, the GMC autocorrelation function in NGC~3351 is unexpectedly less correlated at small scales. This is possibly due to the GMCs which populate the ring of NGC~3351 which may be more uniformly distributed radially around the ring. 

\subsection{Cross-correlation Functions}
\label{sec:rd:ccf}

\begin{figure*}
    \centering
    \includegraphics[width=\textwidth]{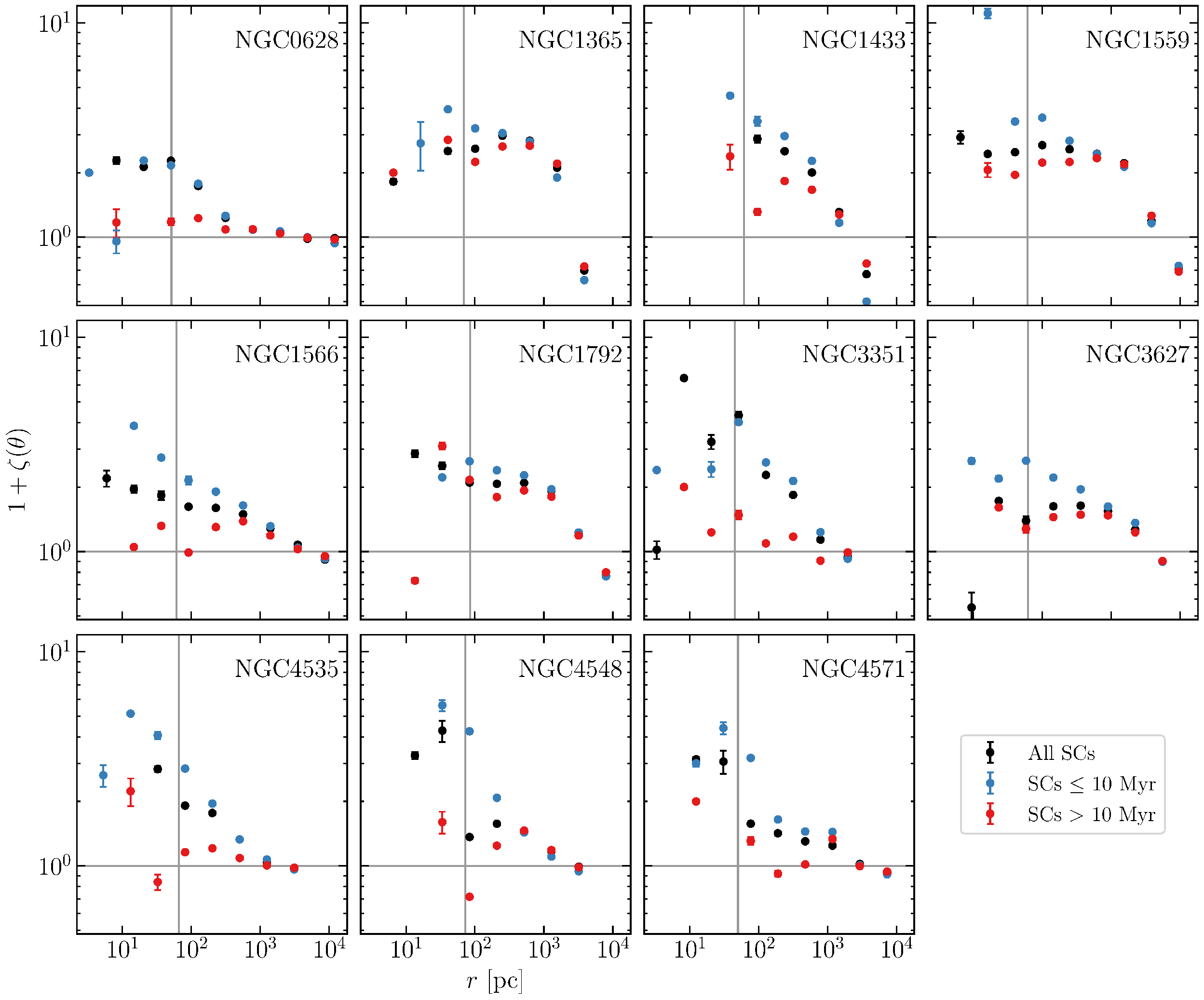}
    \caption{Cross-correlation functions for the GMCs with all star clusters (black), clusters ${\leq}10$~Myr (blue), and clusters ${>}10$~Myr (red) for each galaxy. The horizontal grey line marks a uniform, random distribution at $1 + \zeta(\theta) = 1$. Twice the median radius of the GMCs in each galaxy is shown as the vertical grey line.} 
    \label{fig:cross_grid}
\end{figure*}

We quantify the cross-correlation between the star clusters and GMCs using the methodology given in Section~\ref{subsec:anl:cross}. Results are shown in Figure~\ref{fig:cross_grid}. The cross-correlation functions prove to be difficult to interpret but, in most cases, the young star clusters spatially correlate with GMCs at small spatial scales and the older population is less correlated.  Over time, the fractal distribution is dissipated as star clusters migrate away from their natal GMCs and the GMCs are disrupted. In some of the cross-correlation functions, there is a ``gap'' resembling a piece-wise function. This is because there are no star cluster--GMC pairs found with separations that fall in that bin. 

As mentioned in Section~\ref{sec:rd:acf}, the GMCs are not as highly autocorrelated as the stellar clusters. This may also be affecting the measured cross-correlation functions by lessening the cross-correlation with the young clusters.

\section{Conclusions}
\label{sec:conc}

We present the results of a study combining the PHANGS--ALMA GMC catalogs with the star cluster catalogue from PHANGS--HST for a sample of 11~nearby star-forming galaxies. We spatially correlate the two catalogs first with a simple nearest neighbour analysis which reveals that star clusters with ages ${\leq}10$~Myr are found to lie closer to GMCs on average than the older cluster populations. This follows the trend expected for hierarchical star formation where the clusters inherit the hierarchical distribution of the interstellar medium. Next, we expand the analysis by including information on the sizes of the GMC. We look for a line-of-sight alignment of clusters with GMCs and determine if a cluster is within the radius of the nearest GMC, between 1 and 2~radii of the GMC, between 2 and 3~radii, or beyond which we consider unassociated. To minimize chance alignment, we apply a $10$~km~s$^{-1}$ velocity cutoff. Clusters which are closely associated with a GMC (within the GMC radius) are found to be very young with a median age of $1$~Myr for most of the galaxies. After $\sim6$~Myr, the clusters are no longer associated with their natal gas clouds. This timescale is a measure of the time it takes to dissipate the gas cloud after the onset of star formation. Our analysis follows similar methodology as \citet{grasha19} and we find good agreement with their results in M51. These results also serve as an independent confirmation of the feedback timescale measured using the ``uncertainty principle of star formation'' in \citet{chevance20}.  Specifically, combining the ${\sim}4$~Myr timescale for embedded star formation \citep{kim21} with the ${\sim}3$~Myr timescale for the exposed or unembedded phase of star formation \citep{chevance20} yields a total of ${\sim}7$~Myr, consistent with our finding of $\sim6$~Myr for age until dispersal.

We perform this same analysis broken down by galactic environment and find the same trend where clusters closely associated with GMCs are ${\sim}1$~Myr, and beyond ${\sim}5$~Myr clusters are no longer associated with GMCs. The unassociated clusters found in the centres of the galaxies are measured to be older than $300$~Myr on average. This suggests these clusters could be old globular clusters which reside in the stellar bulges. \citet{turner21} note that there are likely globular clusters in the centre of NGC~3351 but the SED fitting performed on the PHANGS--HST clusters does not recover ages as old as expected (${\sim}10$~Gyr) for these globular clusters, due to the limiting assumption of solar metallicity for all sources. The median age for all clusters within spiral arms is $9$~Myr, while clusters in interarm regions are ${\sim}32$~Myr old. This suggests it takes on average $\sim20$~Myr for a cluster to migrate out of a spiral arm, or that many young clusters are disrupted within ${\sim}20$~Myr.

Autocorrelation functions are measured in Section~\ref{sec:rd:acf} and show the young clusters are more highly autocorrelated at small spatial scales compared to the older cluster populations. We find the GMCs to be nearly uniformly distributed across spatial scales. Power-law fits show that galaxies with distinct spiral arms and rings---NGC~1566, NGC~3351, and NGC~4535---have the largest difference in autocorrelation function slopes between the young and older clusters. We also measure the cross-correlation functions and find that the young clusters track well with the GMCs. However, the cross-correlation functions are difficult to interpret given poor statistics in a few of the galaxies.  Our interpretation is that stellar clusters form at the density peaks of the hierarchy, and are thus likely more strongly clustered than all levels of the hierarchy. But the overall picture is modulated by the sensitivity to galactic morphological properties, which drives global correlations that impact the level of correlation between stellar clusters and GMCs. 

We plan to expand upon the analysis described in this study as more PHANGS--HST star cluster catalogs are produced for more galaxies. This expansion will allow us to build out the statistics for more galaxies and therefore more star clusters. Additionally, a machine learning algorithm is being developed to aid in the identification and classification of PHANGS--HST star clusters \citep{wei20,thilker21}. This will greatly increase the number of identified star clusters within each galaxy, allowing for much better statistical analysis. Finally, a stellar association catalogue for PHANGS--HST is being developed (detailed in Larson et al., in preparation) which provides a better way of identifying the youngest stellar populations which are not captured in the traditional cluster catalog. Performing the analysis presented here on the stellar associations should reveal an even stronger correlation between the young stellar populations and their natal gas clouds. 

\section*{Acknowledgements}
We thank the referee for the helpful recommendations. This work has been carried out as part of the PHANGS collaboration. K.G. is supported by the Australian Research Council through the Discovery Early Career Researcher Award (DECRA) Fellowship DE220100766 funded by the Australian Government. 
K.G. is supported by the Australian Research Council Centre of Excellence for All Sky Astrophysics in 3 Dimensions (ASTRO~3D), through project number CE170100013. TGW acknowledges funding from the European Research Council (ERC) under the European Union’s Horizon 2020 research and innovation programme (grant agreement No. 694343).  JMDK and MC gratefully acknowledge funding from the Deutsche Forschungsgemeinschaft (DFG) through an Emmy Noether Research Group (grant number KR4801/1-1) and the DFG Sachbeihilfe (grant number KR4801/2-1), as well as from the European Research Council (ERC) under the European Union's Horizon 2020 research and innovation programme via the ERC Starting Grant MUSTANG (grant agreement number 714907). FB would like to acknowledge funding from the European Research Council (ERC) under the European Union’s Horizon 2020 research and innovation programme (grant agreement No.726384/Empire). MC gratefully acknowledges funding from the DFG through an Emmy Noether Research Group (grant number CH2137/1-1). This research is based on observations made with the NASA/ESA Hubble Space Telescope obtained from the Space Telescope Science Institute, which is operated by the Association of Universities for Research in Astronomy, Inc., under NASA contract  NAS 5–26555. These observations are associated with program 15654. HAP acknowledges support by the Ministry of Science and Technology of Taiwan under grant 110-2112-M-032-020-MY3. MB gratefully acknowledges support by the ANID BASAL project FB210003 and from the FONDECYT regular grant 1211000. SCOG and RSK acknowledge financial support from DFG via the Collaborative Research Center `The Milky Way System’ (SFB 881, Funding-ID 138713538, subprojects A1, B1, B2, B8), from the Heidelberg Cluster of Excellence `STRUCTURES' (EXC 2181 - 390900948), and from the European Research Council in the ERC Synergy Grant `ECOGAL' (project ID 855130).

This paper makes use of the following ALMA data: ADS/JAO.ALMA\#2015.1.00956.S and ADS/JAO.ALMA\#2017.1.00886.S. ALMA is a partnership of ESO (representing its member states), NSF (USA) and NINS (Japan), together with NRC (Canada), MOST and ASIAA (Taiwan), and KASI (Republic of Korea), in cooperation with the Republic of Chile. The Joint ALMA Observatory is operated by ESO, AUI/NRAO and NAOJ.  The Digitized Sky Survey was produced at the Space Telescope Science Institute under U.S. Government grant NAG W-2166. The images of these surveys are based on photographic data obtained using the Oschin Schmidt Telescope on Palomar Mountain and the UK Schmidt Telescope. The plates were processed into the present compressed digital form with the permission of these institutions. 

Based on observations and archival data obtained with the \textit{Spitzer} Space Telescope, which is operated by the Jet Propulsion Laboratory, California Institute of Technology under a contract with NASA.

\section*{Data Availability}
The GMC catalogue analyzed here is published in \cite{rosolowsky21} and Hughes et al.\ (in preparation).  The stellar cluster catalogue is from \cite{lee22} and is available at the PHANGS homepage at Mikulski Archive for Space Telescopes (MAST) with doi:\href{https://doi.org/10.17909/t9-r08f-dq31}{10.17909/t9-r08f-dq31}. 

\bibliographystyle{mnras}
\bibliography{all}

\bsp	
\label{lastpage}
\end{document}